\documentclass{emulateapj} 
\usepackage{graphicx}
\begin{document} 
 
\title{The Invisibles: A Detection Algorithm to Trace the faintest Milky Way Satellites} \author{S.\ M. Walsh\altaffilmark{1},
B. Willman\altaffilmark{2}, H. Jerjen\altaffilmark{1}}
 
\altaffiltext{1}{Research School of Astronomy and Astrophysics, 
Australian National University, Cotter Road, Weston, ACT 
2611;\email{swalsh@mso.anu.edu.au}} 
\altaffiltext{2}{Clay Fellow, Harvard-Smithsonian Center for Astrophysics, 60 
Garden Street Cambridge, MA 02138}

\begin{abstract} 
 A specialized data mining algorithm has been developed using
  wide-field photometry catalogues, enabling systematic and efficient
  searches for resolved, extremely low surface brightness satellite
  galaxies in the halo of the Milky Way (MW). Tested and calibrated
  with the Sloan Digital Sky Survey Data Release 6 (SDSS-DR6) we
  recover all fifteen MW satellites recently detected in SDSS, six
  known MW/Local Group dSphs in the SDSS footprint, and 19 previously known
  globular and open clusters. In addition, 30 point source
  overdensities have been found that correspond to no cataloged
  objects. The detection efficiencies of the algorithm have been
  carefully quantified by simulating more than three million model
  satellites embedded in star fields typical of those observed in
  SDSS, covering a wide range of parameters including galaxy distance,
  scale-length, luminosity, and Galactic latitude.  We present several parameterizations of these detection limits to facilitate comparison between the observed Milky Way satellite population and predictions. We find
  that all known satellites would be detected with $>90\%$ efficiency
  over all latitudes spanned by DR6 and that the MW satellite census within DR6 is complete to a magnitude limit of $M_V\approx-6.5$ and a distance of 300 kpc. Assuming all existing MW
  satellites contain an appreciable old stellar population and have
  sizes and luminosities comparable to currently known companions, we
  predict lower and upper limit totals of 52 and 340 Milky Way dwarf satellites within $\sim260$ kpc if
  they are uniformly distributed across the sky. This result implies
  that many MW satellites still remain undetected. Identifying and
  studying these elusive satellites in future survey data will be
  fundamental to test the dark matter distribution on kpc scales.
\end{abstract} 
 
%%%%%%%%%%%%%%%%%%%%%%%%%%%%%%%%%%%%%%%%%%%%%%%%%%%%%%%%%%%%%%%% 
\section{Introduction}
The dwarf galaxy population of the Milky Way (MW) provides invaluable
insight into galaxy formation and evolution.  Their resolved stars
reveal their formation histories and enable precise measurements of
their structural parameters, ages and metallicities.  These histories
of individual, nearby systems provide a unique approach to studying
the Universe across the cosmic ages. Dwarf galaxies are also the most
numerous type of galaxy in the Universe and are thought to be the
building blocks of larger galaxies. Owing to their low masses, their
properties may be strongly influenced by ionizing radiation in the
early Universe and by the energy released by supernovae.  The impacts
of both of these are weak links in our understanding of structure
formation. Finding and studying nearby dwarfs of the lowest masses
and luminosities is thus an essential component to understanding
galaxy formation on all scales.

The Milky Way dwarf galaxy population is also at present the most
direct tracer of the abundance, mass spectrum, characteristic size,
and spatial distribution of dark matter on sub-galactic
scales. Standard $\Lambda$CDM simulations of MW-size dark matter
haloes predict many more dark matter sub-haloes than are observed as
dwarf galaxies \citep{Klypin99,Moore99}. The recent ``Via Lactea''
simulation contains 2,000 dark matter sub-halos within 289 kpc of
the simulated primary galaxy \citep{vialactea} which have no observed
optically luminous counterparts. This discrepancy gives rise to the
questions how and in what mass regime do the baryons disappear from
dark matter clumps? Studies of the spatial distributions of Milky Way
and M31 dwarf galaxy companions have also highlighted possible
discrepancies between $\Lambda$CDM theory and observations
\citep{Kroupa05,Kang05,Metz07}.
 
The most obvious reason for these apparent discrepancies in the number
and spatial distributions of dwarf galaxies is substantial incompleteness as the Milky Way halo
has not yet been uniformly searched for dwarf galaxy companions to low
enough luminosities and star densities, in particular close to the
Galactic plane where foreground contamination is severe.  For example,
\cite{Willman04} compared the spatial distribution of MW satellites to
that of M31's population, as well as that of a simulated dark matter
halo and concluded that some dwarfs may have been missed at low
Galactic latitudes, and that the total number of MW satellites with
properties similar to the known objects could be as many as triple the
known population.

The viability of this solution, at least in part, has been underscored
by the recent discoveries of fourteen new Galactic companions from the
photometric data of the Sloan Digital Sky Survey (SDSS). These objects
all appear to be dominated by old ($\gtrsim10$ Gyr) stellar
populations, with the exception of Leo T \citep{leotpop}. Nine of
these companions were immediately identified as dwarf spheroidal (dSph)
galaxies: Ursa Major, Canes Venatici, Bo\"{o}tes, Ursa Major II, Canes
Venatici II, Hercules, Leo IV, Coma Berenices, Leo T and Leo V
\citep{uma,cvn,bootes,uma2,quintet,leot,leov}. Spectroscopic follow-up has
confirmed that they all are highly dark matter dominated dwarf
galaxies \citep{simon07,spectro}.  Willman 1, Segue 1 and Bo\"otes II
\citep{willman1,quintet,boo2} occupy a region of size-luminosity space
not previously known to be inhabited by old stellar populations,
at the intersection of MW dSphs and globular clusters. Spectroscopic
studies \citep{spectro} showed that Willman 1 may be resident inside a
dark matter subhalo with a mass-to-light ratio of $\sim470$. If these
ambiguous objects are gravitationally bound, then tidal arguments also
favor them being dark matter dominated \citep{boo2mmt}. The remaining
two objects discovered in SDSS, Koposov 1 and 2, \citep{k12} are extremely faint Galactic
globular clusters.
 
Numerous authors have shown that the predictions of $\Lambda$CDM
simulations can be reconciled with the small number of observed MW
dwarf galaxies if simple models for baryonic physical processes are
taken into account when interpreting the results of numerical
simulations (e.g. \citealt{bullock01,tumultuous,simon07}).  For
example, \citet{strigari07} show that the central masses ($M_{0.6
  kpc}$) of the Milky Way dwarf galaxies are well-constrained by the
data and that their mass function closely matches the $M_{0.6 kpc}$
mass function of both the earliest forming sub-halos and the most
massive accreted sub-halos in the Via Lactea simulation.
 
A well-defined, deep survey of Milky Way dwarf galaxies over a large
fraction of the sky is critical to assess any of the above scenarios.
The dwarf galaxies detected (or not) by such a survey will provide one
of the best ways to rigorously test the $\Lambda$CDM paradigm by
comparing a variety of metrics (distribution, mass, scale sizes, and
number) of the Milky Way dwarfs with the predictions of $\Lambda$CDM +
galaxy formation models. Willman et al (2002) and \cite{koplf}
have previously conducted automated searches for Milky Way dwarfs in
the SDSS and their corresponding detection limits.  The original
Willman et al survey was only performed over a couple of thousand
square degrees of sky.  The \cite{koplf} survey was performed with a
more sensitive algorithm (critical, because they found many new
satellites to be on the edge of detectability), but few galaxies were
used to accurately quantify their detection limits.

In this paper, we present critical improvements to the present
characterization of the detectability of Milky Way dwarf galaxies over
the $\sim$9,500 square degrees of the SDSS in Data Release 6 (DR6).  We also
present an improved detection algorithm over previous searches.  We
aim to construct the most complete, well-defined census of MW
satellites by embarking on a Milky Way all sky satellite hunt.  This
search will ultimately combine SDSS Data Release 6 (\citealt{dr6}), The 2-Micron All Sky Survey (2MASS; \citealt{2mass}) and
the upcoming Southern Sky Survey \citep{SkyMapper}.  The Southern Sky
Survey will cover the entire 20,000 deg$^2$ below $\delta<0^\circ$ using
the new Australian National University (ANU) SkyMapper Telescope equipped with a 5.7 sq degree wide-
field camera that is currently under construction with survey operation
expected to commence early 2009.
%%%%%%%%%%%%%%%%%%%%%%%%%%%%%%%%%%%%%%%%%%%%%%%%%%%%%%%%%%%%%
\section{SDSS Data}
The Sloan Digital Sky Survey \citep{sdss} is an automated
multi-color imaging and spectroscopic survey spanning 9,500 square
degrees surrounding the North Galactic Pole. The $u,g,r,i$ and $z$
imaging data \citep{sdssfilters,sdsscamera} are photometrically and
astrometrically reduced through an automatic pipeline
\citep{sdssphoto1,sdssphoto2,sdssphoto3,sdssphoto4,sdssastro}. We
subsequently correct for reddening with the \cite{schlegel} extinction
values given in the SDSS catalog. All following work is performed on
point sources from DR6, using the photometry flags from the examples of
database queries appropriate for point sources available on the SDSS
Skyserver website\footnote{http://cas.sdss.org/dr6/}. To ameliorate
effects of incompleteness in the point source catalog and star/galaxy
separation, we only consider sources brighter than $r=22.0$. The
photometric data are provided by the SDSS DR6 \citep{dr6}.  We take
our data from local copies of the SDSS datasets, maintained at the
Harvard-Smithsonian Center for Astrophysics.

%%%%%%%%%%%%%%%%%%%%%%%%%%%%%%%%%%%%%%%%%%%%%%%%%%%%%%%%%%%%%%%
\section{Survey Method}\label{method}
Low surface brightness Milky Way satellites are detectable only by
their resolved stars. With the least luminous known Milky Way
satellites, such as Bo\"otes II, containing fewer than $\sim20$ stars
brighter than the main sequence turn-off (\citealt{boo2mmt}), a deep,
wide-field, uniform, multi-color photometric catalog is essential for
searching for these objects. They will typically reveal their presence
as statistically significant spatial overdensities relative to the
Galactic foreground. Their signal can be enhanced by selecting stellar
sources that are consistent in color-magnitude space with, for
example, an old population of stars at a fixed distance. In this paper, we
restrict ourselves to the old stellar populations characteristic of
Local Group dSphs, but the population-specific elements of the
algorithm can be easily modified for other systems. The strategy of
our detection algorithm is built upon that of \cite{Willman02} and
\cite{Willman03} which utilized the photometric catalogs from SDSS and
led to the discoveries of Ursa Major \citep{uma} and Willman 1
\citep{willman1}. It is also similar in spirit to \cite{quintet} and
\cite{koplf}. Several systematic searches for MW dwarfs have also been done with non-SDSS data \citep{irwinconf,kleynasearch,whitingsearch}.

In summary, our algorithm applies color and magnitude cuts to stars
in the DR6 catalog, stores their distribution in a spatial array with
$0.02^{\circ}\times0.02^{\circ}$ pixels, spatially smooths the array
with a Plummer surface density profile, and sets comprehensive
thresholds for detection. Each of these steps is described in detail
in the following sections.
 
\subsection{Data Management}
In order to efficiently manage thousands of square degrees of survey data in
a catalog containing tens of millions of stars, we first divide the
dataset (in the case discussed in this paper, SDSS DR6) into stripes,
each spanning $3^{\circ}$ in declination (to avoid projection effects)
with $2^{\circ}$ of overlap in Dec between adjacent stripes. This
overlap creates a substantial redundancy to ensure that real objects
are situated in the central $\sim2^{\circ}$ of Declination in at least
one stripe, away from possible edge-effects introduced at the stripe
boundaries during the processing described in \S3.3. We then take the
longest continuous regions of the DR6 footprint in RA.

\subsection{Selection Criteria} \label{cuts} 
The mainly old,
metal-poor stars of a nearby dwarf galaxy will occupy a well defined
locus in the color-magnitude diagram (CMD), in contrast to Milky Way
stars which span a wide range in distance, age, and
metallicity. Therefore selecting stars that are consistent in
color-magnitude space with a population of old stars at a particular
distance will significantly enhance the clustering contrast of a dwarf
galaxy's stars over the foreground noise from Milky Way stars.

\begin{figure}[!ht]
 \includegraphics{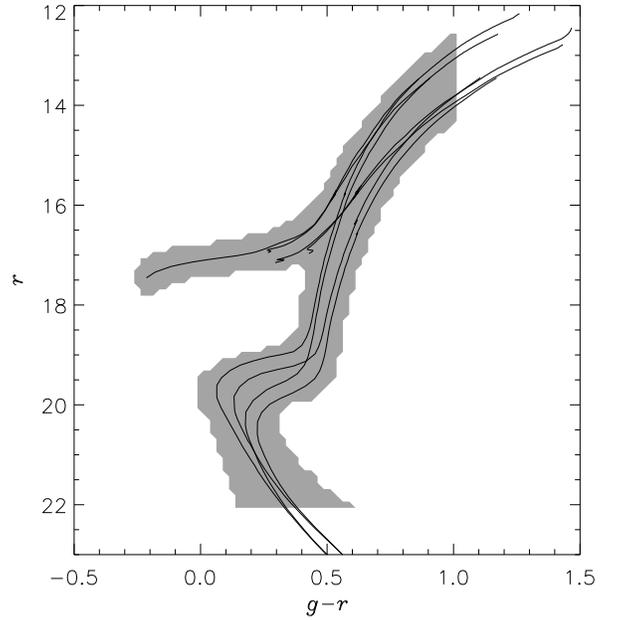}
\caption{$(r,g-r)$ color-magnitude diagram showing the two reddest and
two bluest theoretical isochrones for old stellar populations
([Fe/H]$=-2.27$, $-1.5$ and age 8, 14 Gyr) at a distance modulus of
$m-M=16.5$ ($\sim20$ kpc), generated from Girardi et al. (2004). The shaded region shows pixels that pass the selection criteria.} \label{fig:temp}
\end{figure}

  We use theoretical isochrones in SDSS filters from \cite{sdssiso} to
  define the regions of $(r,g-r)$ space likely to be populated by
  old, metal-poor stars.  \cite{simon07} obtained spectra of stars in
  eight of the newly discovered dwarfs: CVn, CVn II, Com, Her, Leo IV,
  Leo T, UMa and UMa II and found mean abundances in the range
  $-2.29<$[Fe/H]$<-1.97$. Based on this result, we consider isochrones
  for populations with abundances of [Fe/H] = $-1.5$ and $-2.27$ (the
  lower limit in \citealt{sdssiso}) and with ages 8 and 14 Gyr. Four
  isochrones in these ranges can be used to bound the region of CMD
  space we are interested in, namely the four combinations of [Fe/H] =
  $-1.5$ and $-2.27$ and ages 8 and 14 Gyr. Figure \ref{fig:temp}
  shows these four isochrones projected to a distance of 20 kpc.

  We define the selection criteria by the CMD envelope inclusive of
  these isochrones +/- the 1 sigma $(g-r)$ color measurement error as
  a function of $r$ magnitude. Shifting these isochrones over
  distances between $m-M$ = 16.5 -- 24.0 in 0.5 mag steps defines 16
  different selection criteria appropriate for old stellar populations
  between $d\sim20$ kpc and $\gtrsim630$ kpc. We truncate our
  color-magnitude selection template at a faint magnitude limit of
  $r=22.0$, beyond which photometric uncertainties in the colors and
  star/galaxy separation limit the ability to detect these
  populations. We also truncate the selection template at $g-r=1.0$ as
  including redder objects adds more noise from Milky Way dwarf stars
  than signal from more distant red giant branch (RGB) stars. Finally
  we do not include stars with $\delta g$ or $\delta r$ $>$ 0.3 mag in
  our analysis.  To efficiently select stars within this CMD envelope,
  we treat the CMD as an image of $0.025\times0.125$ (color $\times$
  mag) pixels and determine which stars fall into pixels classified as
  ``good'' according to the selection criteria.  Figure \ref{fig:temp}
  shows an example of the selection criteria, in this case for
  $m-M=16.5$ ($\sim20$ kpc). The shaded region highlights pixels that
  would be classed as ``good'' for a system at $\sim20$ kpc.
 
\subsection{Spatial Smoothing}\label{smooth}
After the photometric cuts are applied, we bin the spatial positions
of the selected stars into an array, $E$, with
$0.02^{\circ}\times0.02^{\circ}$ pixel size. We then convolve this 2D
array with a spatial kernel corresponding to the expected surface
density profile of a dSph.  We refer to this smoothed spatial array as
$A$. For our spatial kernel we use a Plummer profile with a $4.5'$
scale-length. This value provides an effective compromise between the
angular scale-lengths of compact and/or distant objects with those of
closer/more extended objects. For reference the angular sizes of the
new satellites are listed in Table \ref{tbl:ang}. We use the $r_h$
values derived by \cite{martinhood} except for Leo V \citep{leov}.

\begin{deluxetable}{lr}
\tablecaption{Angular sizes of the satellites detected in SDSS.\label{tbl:ang}} \tablewidth{0pt} \tablehead{
\colhead{Object}  & \colhead{$r_{h}$} \\
 & \colhead{(arcmin)} }
 \startdata
Bo\"otes         &  12.6 \\
Bo\"otes II         &  4.2    \\
Canes Venatici      &  8.9     \\
Canes Venatici II     &  1.6     \\
Coma Berenices      &  6.0     \\
Hercules           &  8.6     \\
Leo IV              &  2.5    \\
Leo V			& 0.8 \\
Leo T               &  1.4     \\
Segue 1           &  4.4    \\
Ursa Major           &  11.3    \\
Ursa Major II          &  16.0    \\
Willman 1           &  2.3    \\
 \enddata
\end{deluxetable}

The normalized signal in each pixel of $A$, denoted by $S$, gives the
number of standard deviations above the local mean for each element:
\begin{eqnarray*} S =
\frac{A-\bar{A}}{A_{\sigma}}. \nonumber
\end{eqnarray*}
The arrays of running means, $\bar{A}$, and running standard deviations,
$A_{\sigma}$ are both calculated over a $0.9^{\circ}\times0.9^{\circ}$
window around each pixel of $A$.  In particular, $A_{\sigma}$ is given by:
\begin{eqnarray*}
A_{\sigma} = \sqrt{\frac{n (A-\bar{A})^{2}\ast B - ((A-\bar{A}) \ast B)^{2}}{n(n-1)}}.
\end{eqnarray*}
$B$ is a box-filter with $n$ elements and is the same size as the
running average window. The resulting array $A_{\sigma}$ gives the
standard deviation value for each pixel of $A$ as measured over the
$0.9^{\circ}\times0.9^{\circ}$ span of the filter. In the next
section, we will define the detection threshold of this survey in
terms of $S$, as well as in terms of the local stellar density $E$.

\subsection{Detection Threshold(s)}\label{threshold}
In a large survey such as ours, it is critical to set detection
thresholds strict enough to eliminate false detections but loose
enough to retain known objects and promising candidates. To
characterize the frequency and magnitude of truly random fluctuations
in stellar density analyzed with our algorithm, we measure the maximum
value of $S$ for 199,000 $5.5^{\circ}\times3^{\circ}$ simulated fields
of randomly distributed stars that have been smoothed as described in
the previous section. The only difference is that there is no gradient
in stellar density across each field. In the interest of computational
efficiency we do not use a running window for the mean and $\sigma$ of
each simulated field. The field size is chosen such that 1,000 fields
roughly totals an area equal to the DR6 footprint (neglecting regions
lost during convolution). We select 199 stellar densities $n_*$ to
simulate, linearly spaced between 10 and 4,000 stars per square
degree. This range of stellar densities is to model the density range
we find \emph{after} applying the color-magnitude selection criteria
described in \S \ref{cuts} across the SDSS.  In \S \ref{fields} we
study the variation of detection limits with Galactic latitude
(foreground stellar density); the typical number densities we will
consider there are higher than 10 - 4,000 stars per square degree
because we wish to parametrize the detection limits in terms of the
density of \emph{all} stars bluer than $g-r = 1.0$ and brighter than
$r=22$.

\begin{figure}[!ht]
 \includegraphics{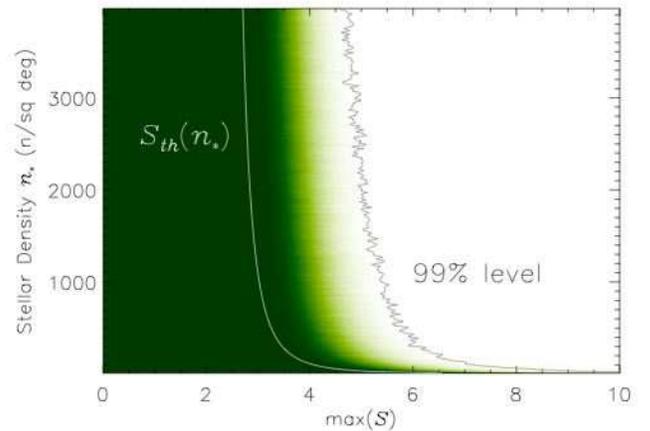}
\caption{A two dimensional cumulative histogram showing the
distribution of max($S$) values for smoothed fields as a function of
different stellar densities $n_*$. The grey line bounds 99\% of the max($S$) values and
the white line shows our threshold density, $S_{th}=f(n_*)$.}
\label{fig:threshold}
\end{figure}

Figure~\ref{fig:threshold} shows a two dimensional cumulative
histogram of the 199,000 max($S$) values as a function of stellar
density.  In low density fields, the distribution of pixel values
becomes non-Gaussian so a simple, global threshold value is
insufficient. The solid grey line shows the contour containing 99\% of the 199,000 max($S$) values at each density. If we simply used a value like this as our threshold, we would
be biasing ourselves against detecting extended objects: Large angular
scale-length systems may not have a peak pixel value above this value,
for example because stars in the object itself increases the local
running mean and sigma. However, such an overdensity may have a
characteristic area larger than any random fluctuation.

\begin{figure}[!ht]
\center
 \includegraphics{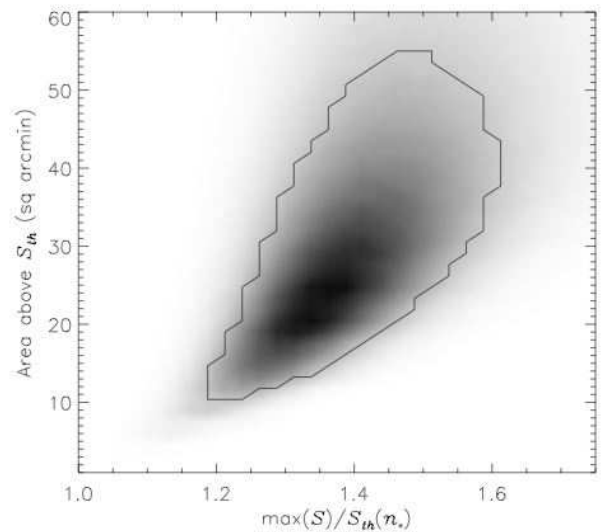}
\caption{``Detection'' area versus max($S$)/$S_{th}(n_*)$
for the 199,000 random fields. The black contour shows the level at
which purely random clustering would produce one false detection over
the approximate area of DR6.} \label{fig:detthings}
\end{figure}

We thus define a detection threshold based on both the peak density
and the characteristic area of an overdensity. To define such an area
we scale down the 99\% contour from Figure \ref{fig:threshold} and
define a threshold density $S_{th}(n_*)$ as a function of stellar
number density. Then, using the 199,000 random fields we examine the
relationship between the peak density max$(S)$ divided by
$S_{th}(n_*)$ and ``detection'' area, i.e. the area of contiguous
pixels of $S$ that have values above $S_{th}(n_*)$ (the white
line). Figure \ref{fig:detthings} shows this area versus
max($S$)/$S_{th}(n_*)$ for the random fields. If we assume a purely
random foreground, we would expect one false detection in the DR6
footprint above an area of $\sim55$ square arcminutes $or$ above a
peak density of $\sim1.6\times S_{th}(n_*)$.  These numbers are set by
the factor by which we scale down the threshold function and are
themselves arbitrary.

Based on the results of these simulations we set the area threshold to
a more conservative 60.0 sq arcmin and the density threshold to a
more conservative $1.75\times S_{th}(n_*)$ to eliminate false positive
fluctuations while preserving all of the known objects within DR6,
including Bo\"otes II \citep{boo2} and the Koposov 1 and 2 globular
clusters \citep{k12}. Thus, a detection is defined as a region where:
 \begin{itemize}
\item{the area of a group of pixels of $S$ contiguously above $S_{th}(n_*)$
(white line, Fig \ref{fig:threshold}) is greater than 60.0 square
arcminutes}\\
\emph{or}
\item{any single pixel value is greater than $1.75\times S_{th}(n_*)$.}
\end{itemize}

We implement these adaptive density thresholds as a function of local
stellar density $n_*$, so that the algorithm may be run over large
fields with varying density and allow direct comparison between fields
of greatly different densities.  The stellar density $n_*$ is
calculated for each pixel of the smoothed, normalized, spatial array
$S$, as the $0.9^{\circ}\times0.9^{\circ}$ running average of the
original spatial density array $E$.

To summarize our algorithm:
  \begin{itemize}
\item{Apply CMD cuts, bin spatial positions of remaining stars into $E$}
\item{Smooth $E$ with Plummer profile to get $A$}
\item{Calculate the $0.9^{\circ}\times0.9^{\circ}$ running mean $\bar{A}$ and running standard deviation $A_{\sigma}$}
\item{Define $S$ as $S=(A-\bar{A})/A_{\sigma}$}
\item{Calculate array of threshold values $S_{th}$ as function of stellar density $n_*$ (from $0.9^{\circ}\times0.9^{\circ}$ running mean of $E$)}
\item{Detections are where contiguous regions of pixels with $S > S_{th}$ is greater than 60.0 sq arcmin \emph{or} any single pixel is greater than $1.75\times S_{th}$.}
\end{itemize}

\subsection{Identifying and Evaluating Detections}
For each of our DR6 data strips defined in \S3.1, the steps outlined
in the previous sections are repeated in 0.5 magnitude distance
modulus intervals, and these 16 frames are layered to form a
3-dimensional array. This 3D approach eliminates complications with
multiple detections of a single object using selection criteria for
different distance moduli, and selects out the strongest detection. The
coordinates of stars within each detection and the CMD within the
detection's area are plotted for later visual inspection. Galaxy
clusters and point sources around partially resolved background
galaxies (such as their associated globular clusters) will contaminate
the detections, but these can be identifiable based on their CMDs (see
Figure \ref{fig:dets4} in \S\ref{dr6}), leaving a list of potential
new Milky Way satellite galaxies and globular clusters. At this point
follow up observations are typically necessary to confirm the
existence and nature of these candidates.

%%%%%%%%%%%%%%%%%%%%%%%%%%%%%%%%%%%%%%%%%%%%%%%%%%%%%%%%%%%%%%%%%%%%%%%
\section{Application to SDSS Data Release 6}\label{dr6}
\begin{figure}[!ht]
\center
\includegraphics{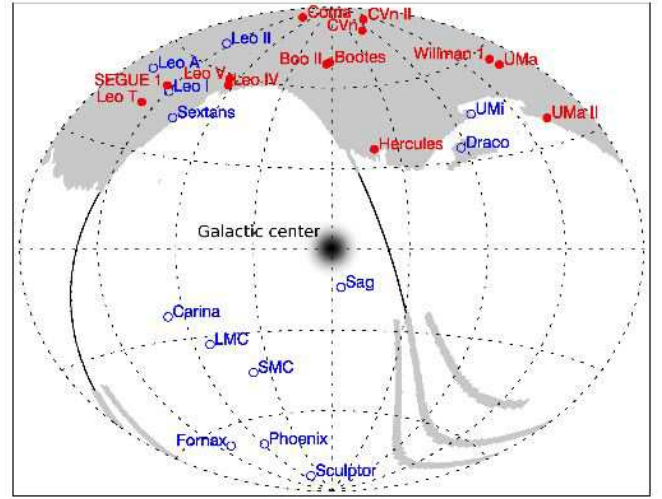}
\caption{Aitoff projection of the DR6 footprint in Galactic
coordinates, centered on the Galactic center. Previously known dwarfs
are marked with open blue circles, satellites discovered in SDSS are
marked with filled red circles.}
 \label{fig:sdssaitoff}
\end{figure}

We apply our search algorithm (as described in \S3) to 21,439,777
sources with $r<22.0$ and $g-r< 1.0$ in the 9,500 deg$^2$ of imaging
data in Data Release 6 of the SDSS.  The DR6 footprint is shown in
Figure \ref{fig:sdssaitoff}, along with previously known dSphs (open
blue circles) and satellites discovered in SDSS (closed red circles).
 
\begin{figure*}[!ht]
\center
 \includegraphics{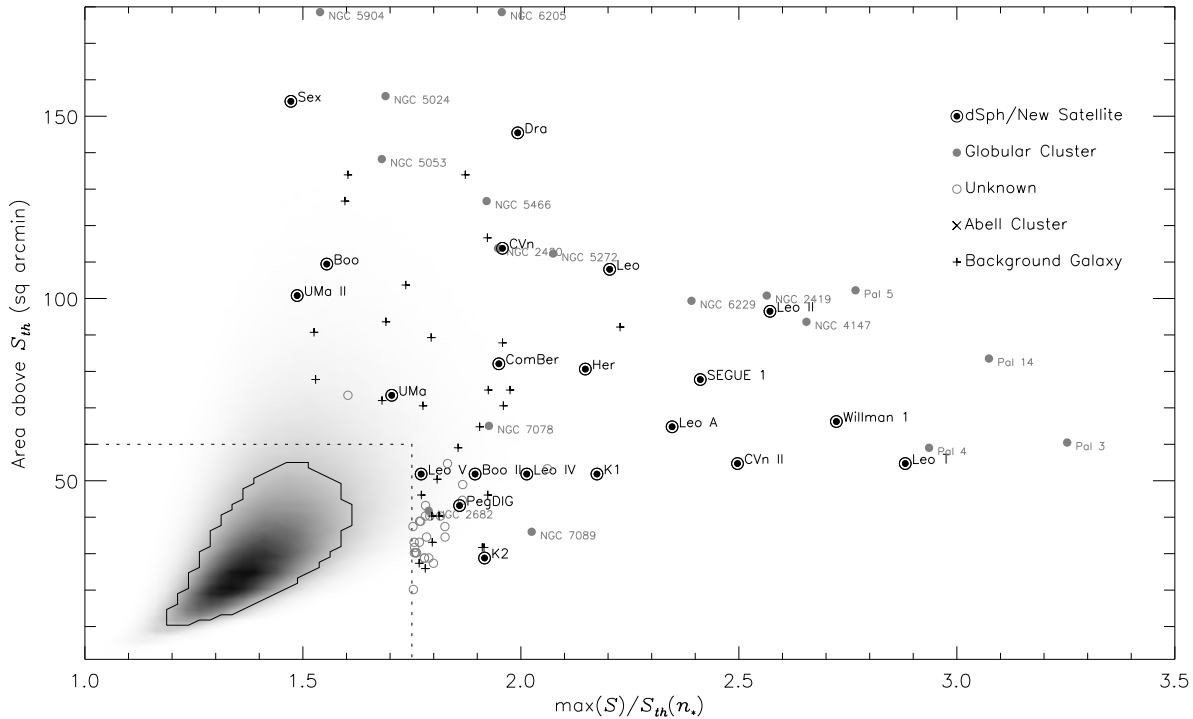}
 \caption{Same as Figure \ref{fig:detthings} but showing all
   detections in DR6. Dotted black lines show the adopted
   thresholds. Galactic/Local Group dSphs and Koposov 1 and 2 are
   shown as black filled circles. The brightest objects such as Draco
   and Sextans increase the average stellar density of the area they
   occupy, which increases $S_{th}(n_*)$ in their vicinity. This means
   that max$(S)/S_{th}(n_*)$ is not as high as one might expect.}
\label{fig:detparam}
\end{figure*}

The significance of our detections of known objects in terms of their
peak density and area are shown in Figure \ref{fig:detparam}. In the
total area of DR6 analyzed, we find 100 unique detections above the
thresholds, defined by the dotted lines of Figure
\ref{fig:detparam}. The positions of each of these detections are
cross-referenced against the SIMBAD database
\footnote{http://simbad.u-strasbg.fr/simbad/} as well as visually
inspected via the SDSS Finding Chart
Tool\footnote{http://cas.sdss.org/astrodr6/en/tools/chart/chart.asp}. Of
our 100 detections, 19 are MW/Local Group dwarfs (counting Bo\"otes
II, Willman 1 and Segue 1), 17 are Galactic globular clusters
(including Koposov 1 and 2), 2 are known open clusters, 28 are
clustering of point sources associated with background galaxies such
as unresolved distant globular clusters, and four are Abell galaxy
clusters. The remaining 30 do not correspond to any catalogued
objects, but color-magnitude diagrams of only a handful of these are
consistent enough with a faint MW satellite to warrant follow-up. The
remainder may be galaxy clusters whose detected center differs from
its cataloged centre by more than $\sim0.25^{\circ}$, or perhaps tidal debris. If the MW stellar halo is the result of accretion of dSph then evidence of this accretion is expected. It should be noted that objects with relatively large
angular size, such as Draco and Sextans, substantially increase the
average stellar density of the area they occupy which increases the
threshold density, meaning they are not as high above the density
threshold as one might expect. Due to the area threshold however, they
are still very prominent detections.
 
\begin{figure*}
 \includegraphics{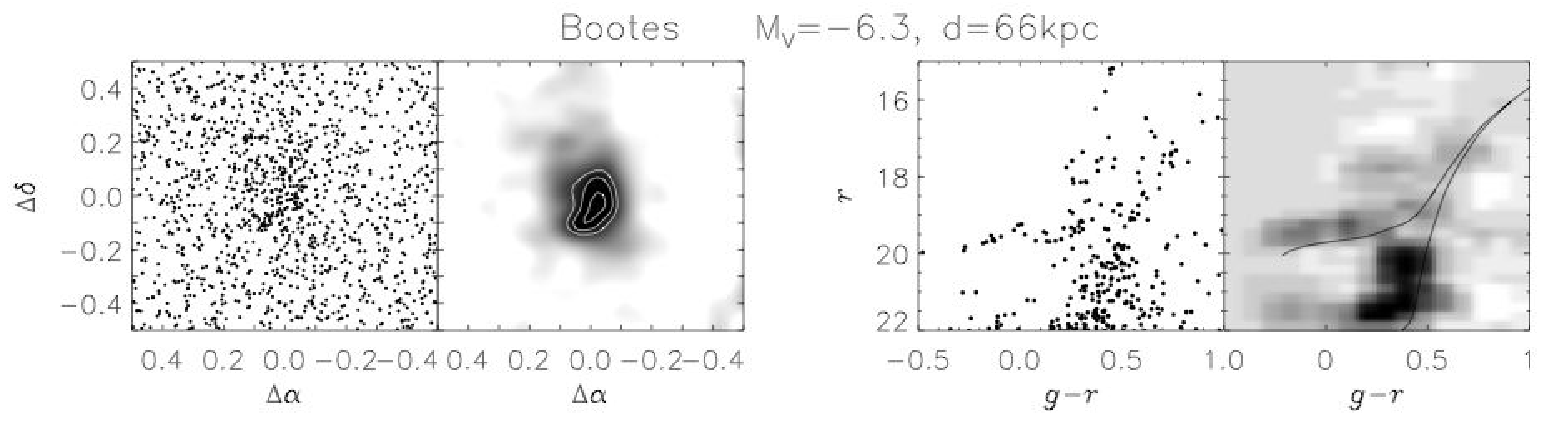}
 \includegraphics{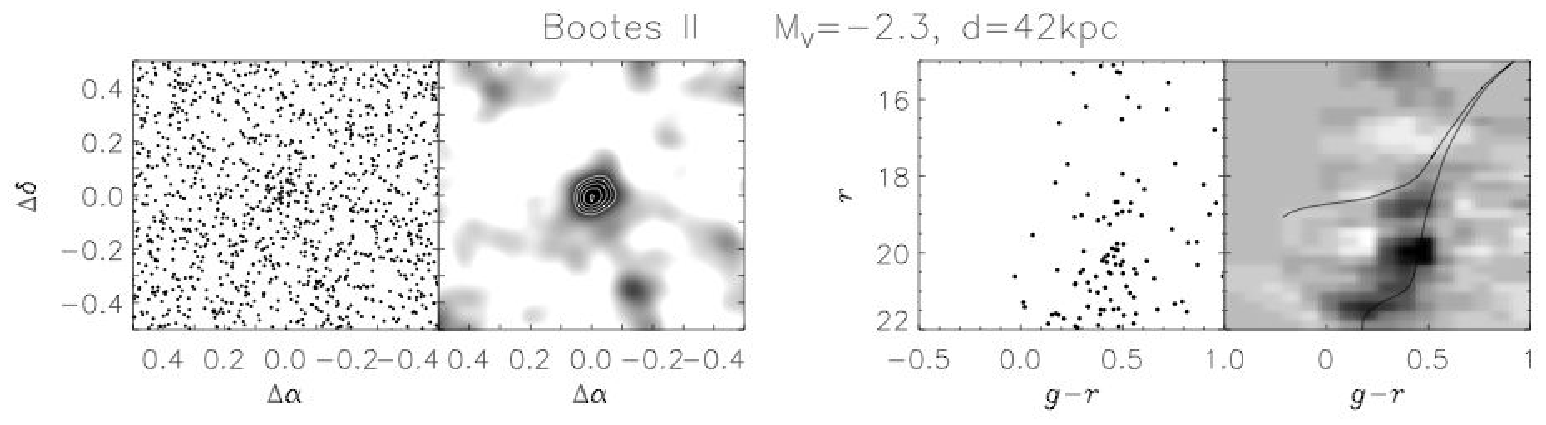}
 \includegraphics{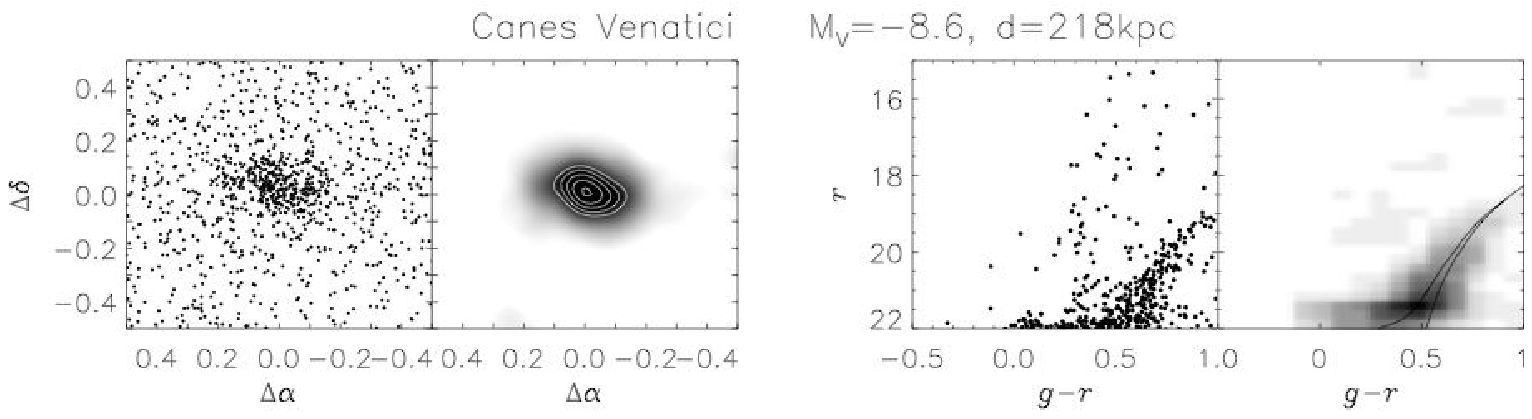}
 \includegraphics{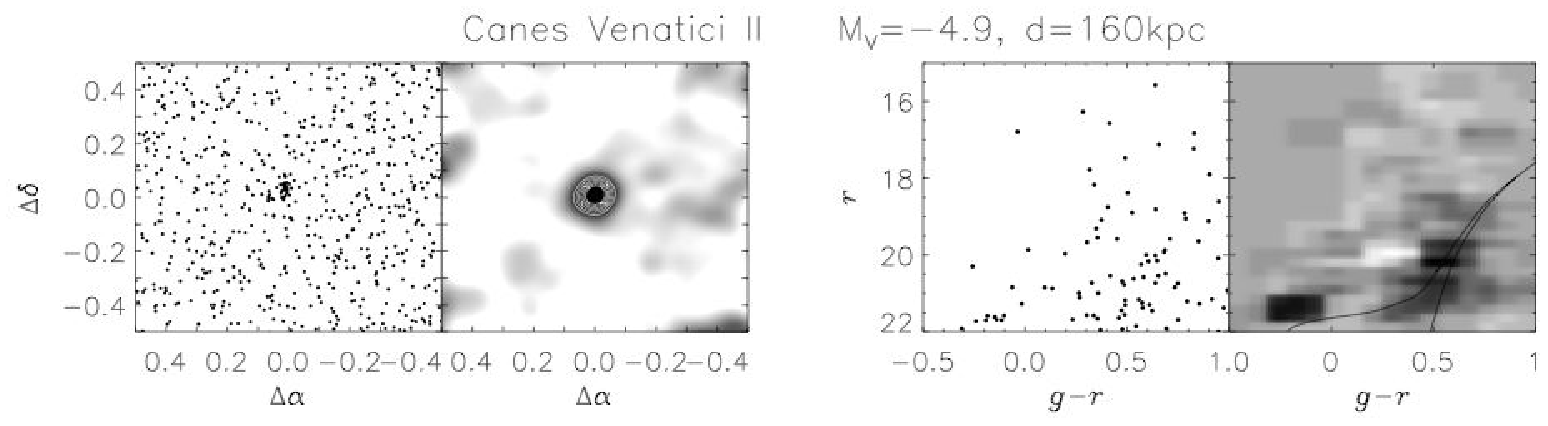}
 \includegraphics{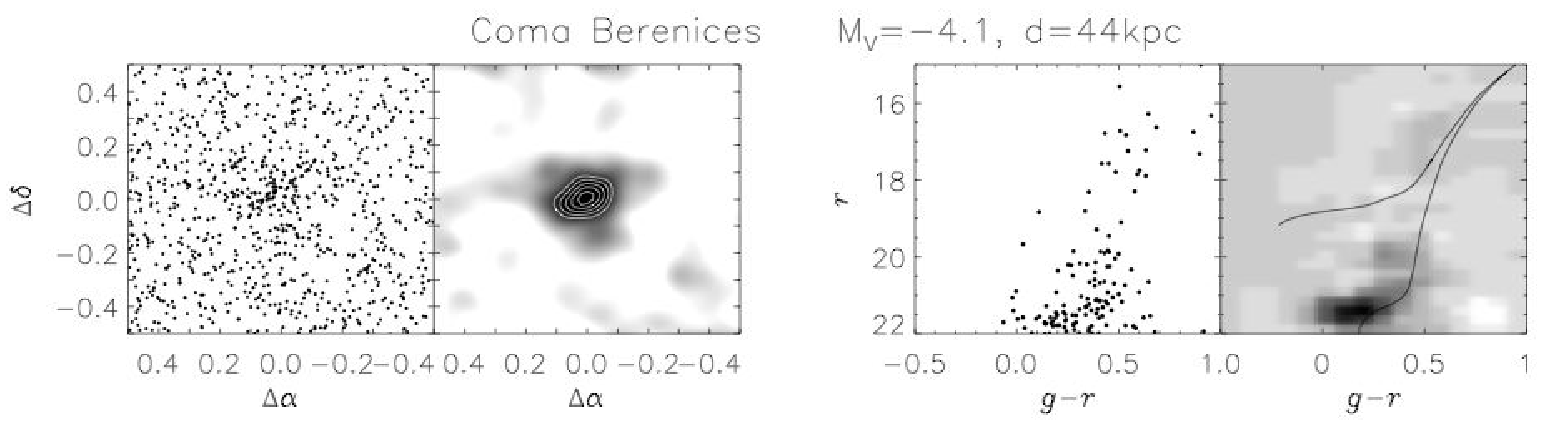}

\caption{Our detections of recently discovered MW
satellites. \emph{Left}: Spatial plot of sources passing selection
cut. \emph{Middle Left}: Contour of smoothed spatial plot. Contours
show 1.0, 1.2, 1.4, 1.6, 1.8, and 2.0 times the density threshold. \emph{Middle
Right}: CMD of region enclosed by contours. \emph{Right}: Hess diagram
of same region, with 13 Gyr, [Fe/H]$=-2.3$ Girardi et al. (2004) isochrone at the objects' distance
overplotted.}
\label{fig:dets}
\end{figure*}

\begin{figure*}
 \includegraphics{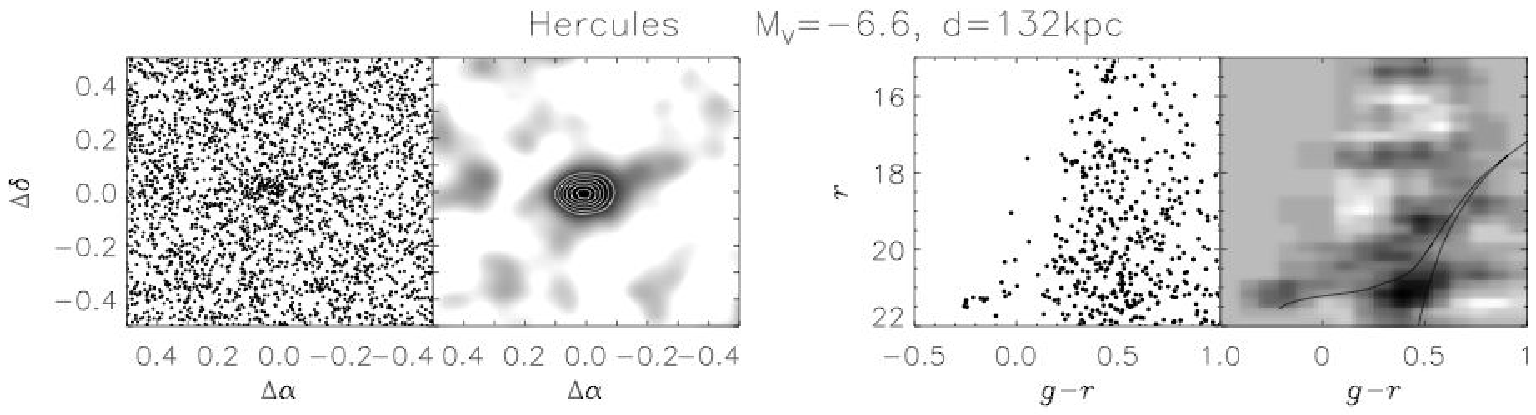}
 \includegraphics{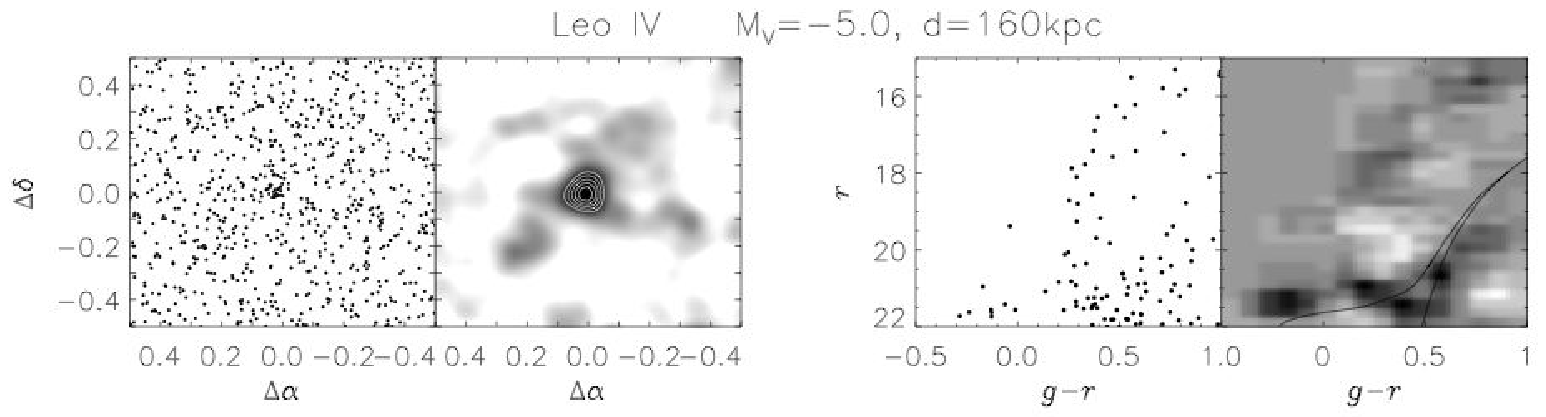}
 \includegraphics{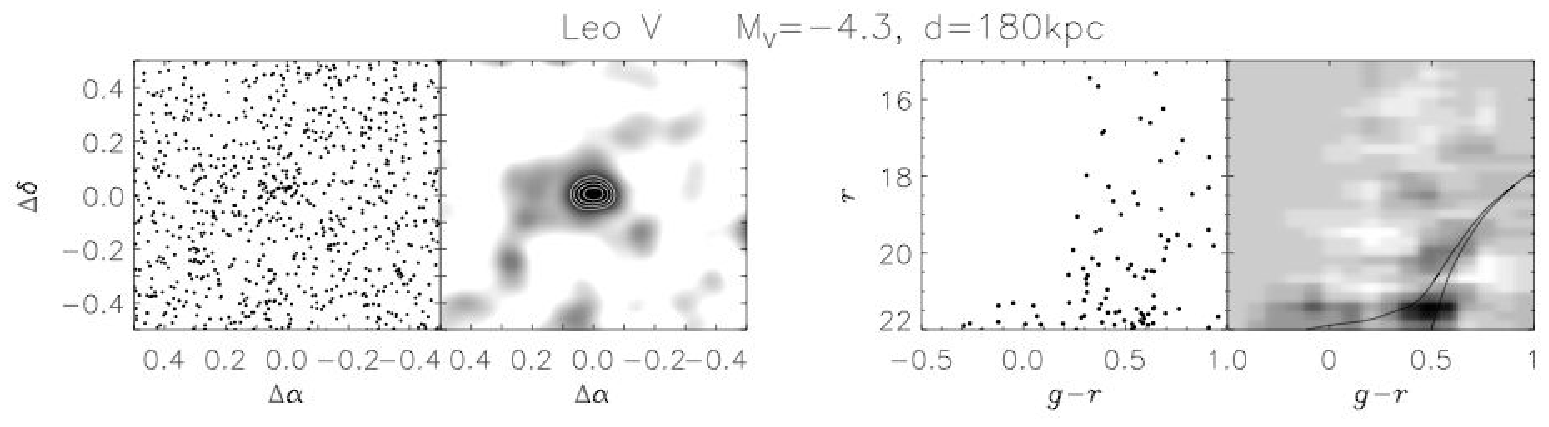}
 \includegraphics{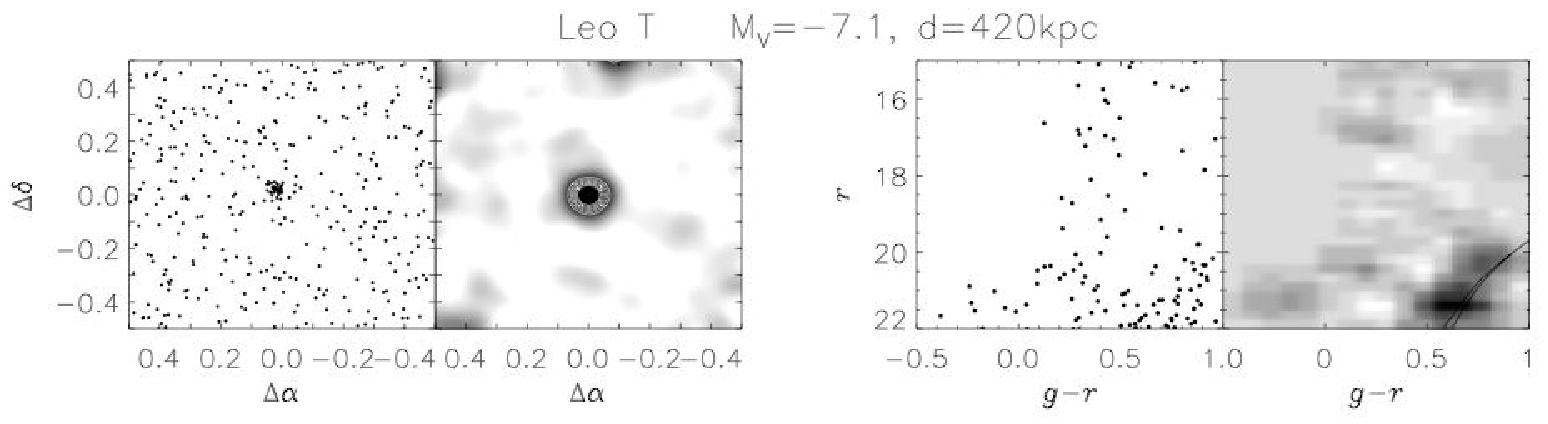}
 \includegraphics{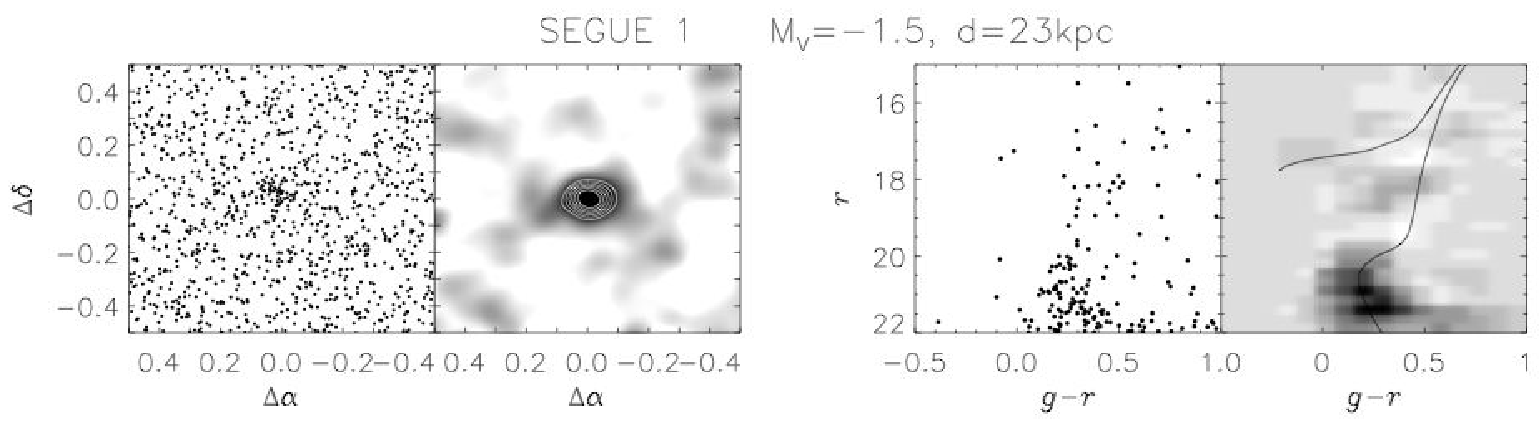}
\caption{Detections of recently discovered MW satellites \emph{cont}.}
\label{fig:dets2}
\end{figure*}

\begin{figure*}[!ht]
 \includegraphics{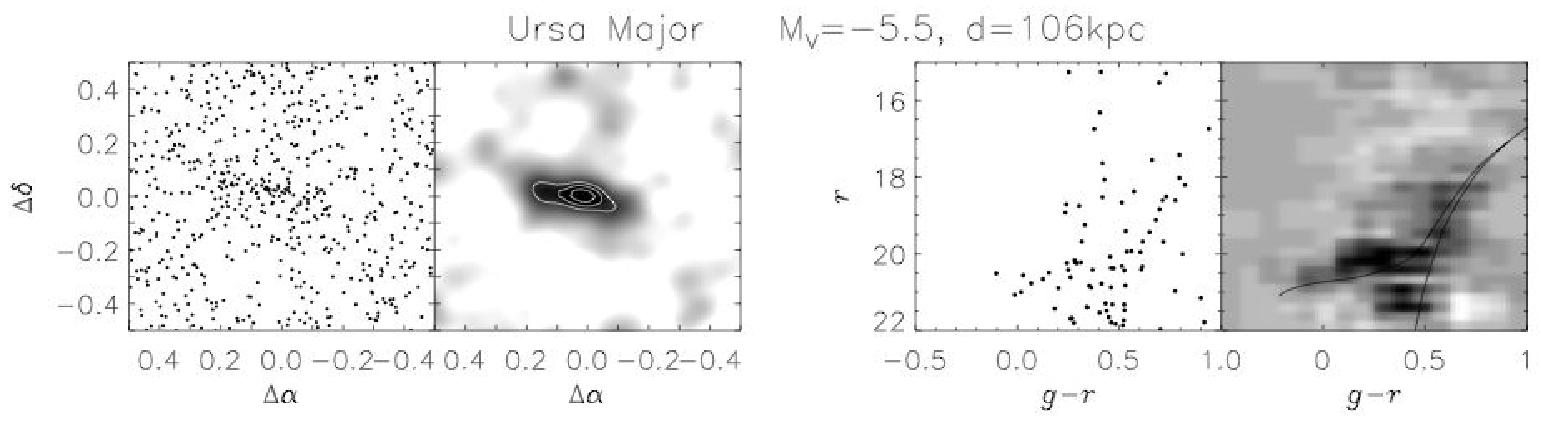}
 \includegraphics{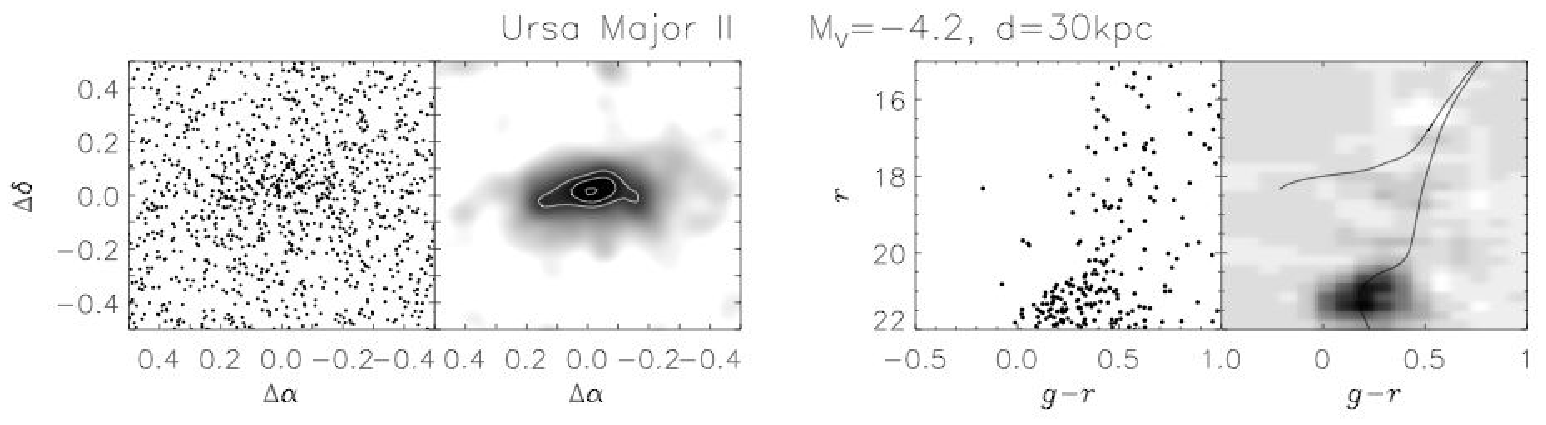}
 \includegraphics{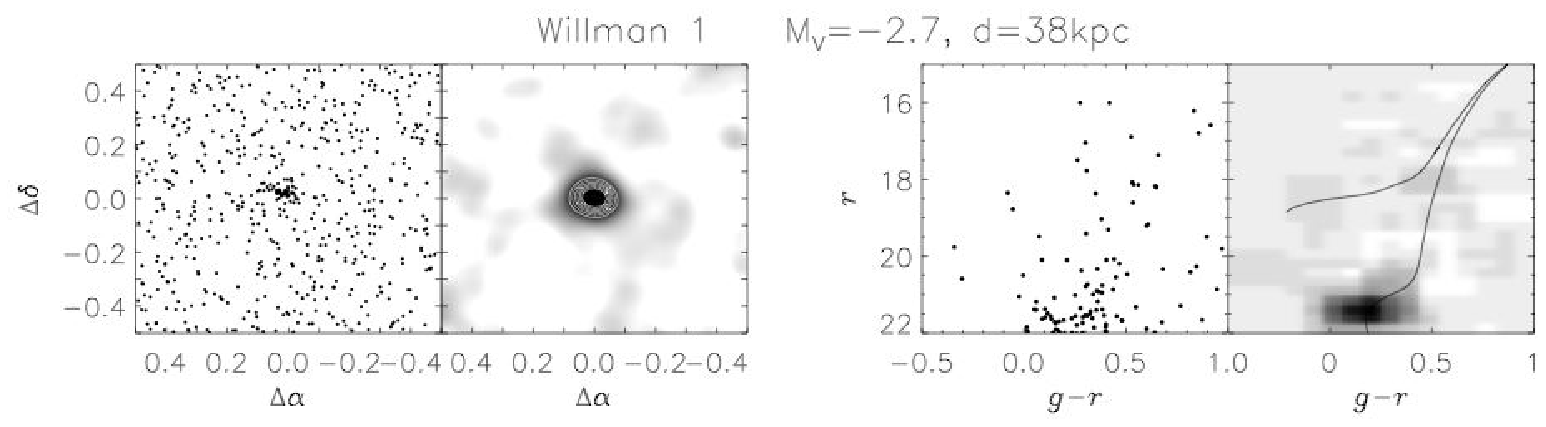}
 \includegraphics{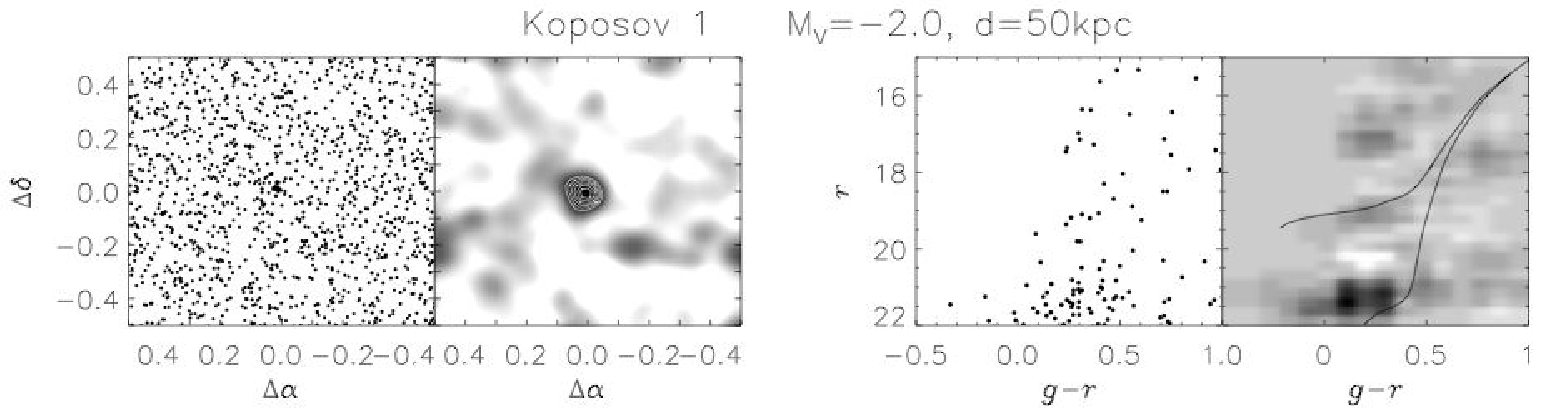}
 \includegraphics{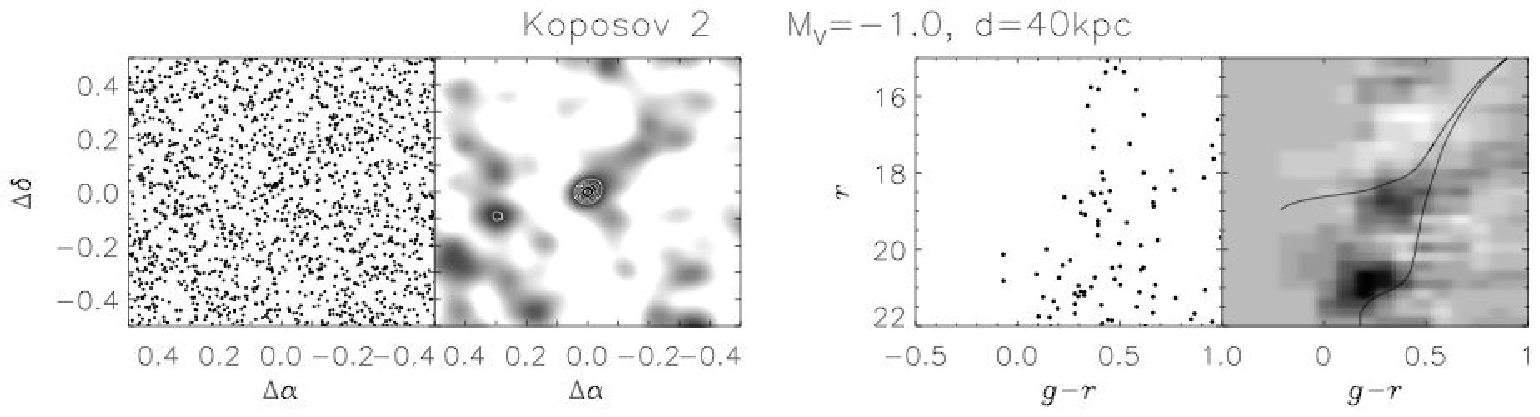}
 
\caption{Detections of recently discovered MW satellites \emph{cont}.}
\label{fig:dets3}
\end{figure*}

We recover all of the newly discovered objects that are within DR6 and
the ``classically'' known Draco, Leo, Leo II, Leo A, Sextans, and
Pegasus DIG dwarfs. Our detections of the new dwarfs are presented in
Figures \ref{fig:dets}, \ref{fig:dets2}, and \ref{fig:dets3}.  These
figures are identical to those output by the automated algorithm for
each detection, aside from the addition of figure titles ($M_V$ and
distances from \citealt{martinhood} and references therein). The left
panel shows the spatial positions of stars passing the photometric selection
criteria at the distance modulus the object was most strongly detected
at. The middle-left panel shows the contour plot corresponding to $S$,
where the contour levels are $(S)/S_{th}(n_*)$ = 1.0, 1.2, 1.4, 1.6,
1.8, and 2.0. The middle-right panel is the CMD of the detection area
and the right panel is the field subtracted Hess diagram. The
isochrone is that of a 13 Gyr, [Fe/H]$=-2.3$ from \cite{sdssiso} at
the distance specified. Besides demonstrating the effectiveness of our
algorithm, these detections provide a benchmark with which to compare
candidates and determine which are consistent with being a new dwarf
satellite.
 
\begin{figure*}[!ht] 
 \includegraphics{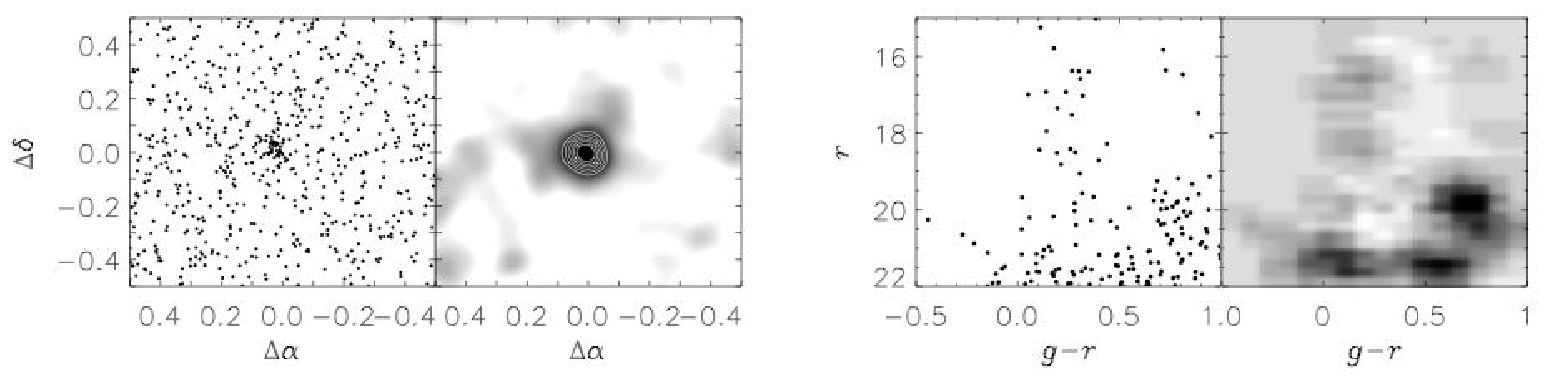} 
 \includegraphics{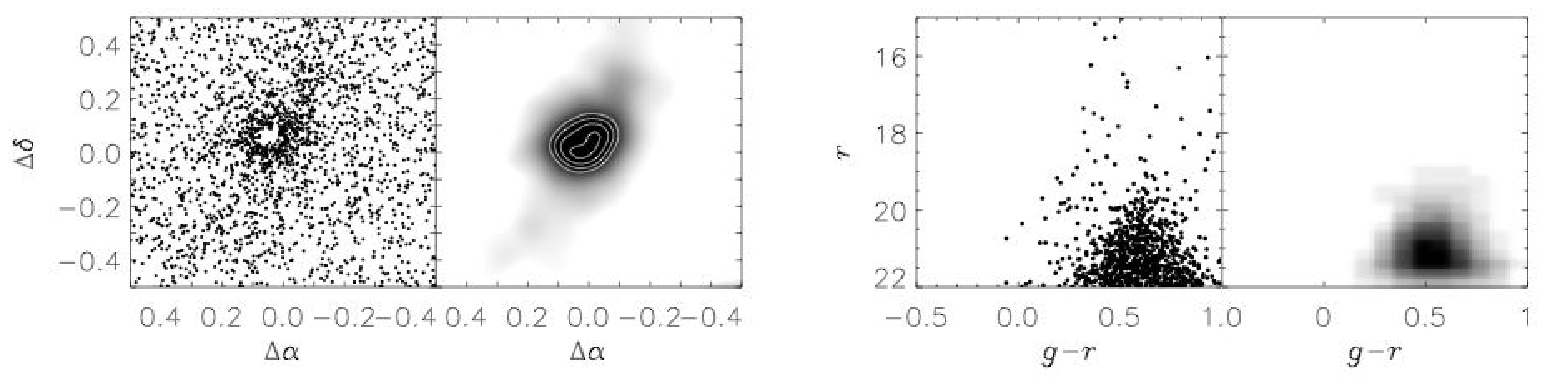} 
\caption{Detections of Abell 1413 (top) and NGC 4486 (bottom) as examples of galaxy cluster and background galaxy detections.} 
\label{fig:dets4} 
\end{figure*}

To further illustrate the product of our algorithm we also show
examples of undesired detections in Figure \ref{fig:dets4}: the galaxy
cluster Abell 1413 (top) and Virgo cluster galaxy NGC 4486
(bottom). These represent typical detections of background galaxies
and galaxy clusters.  

\subsection{Candidate Milky Way Satellites}
\begin{figure*}[!ht] 
 \includegraphics{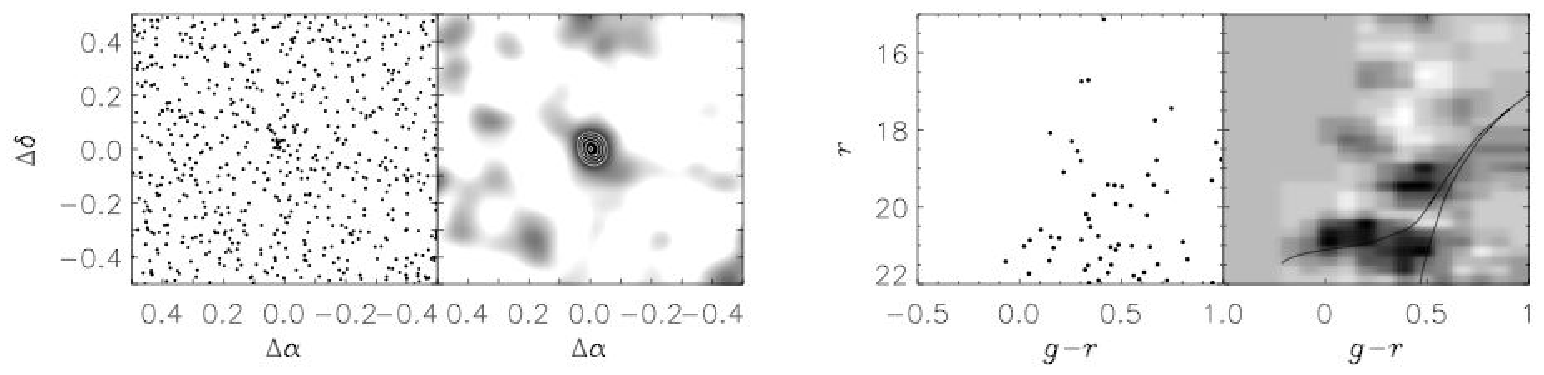} 
 \includegraphics{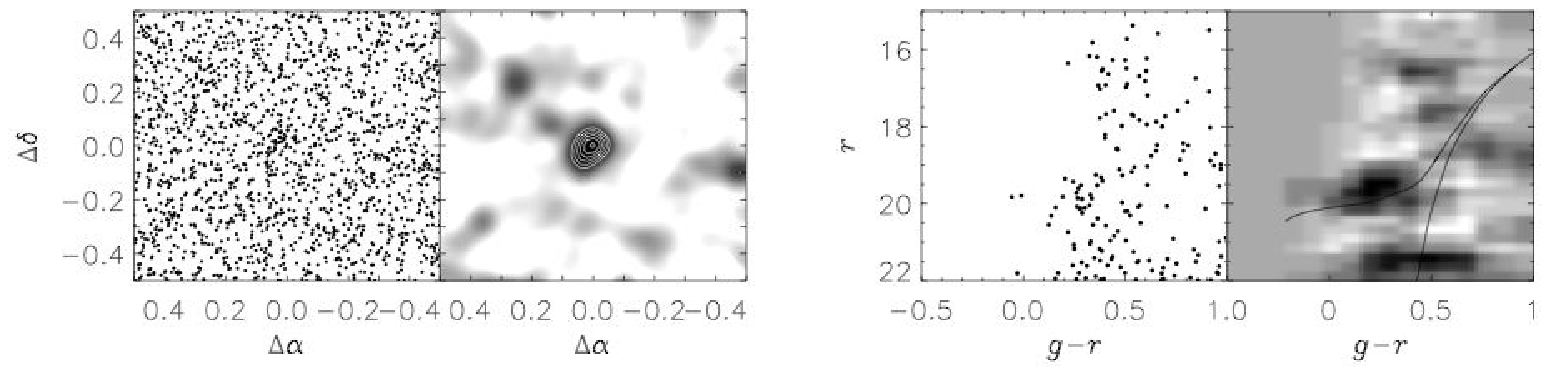} 
 \includegraphics{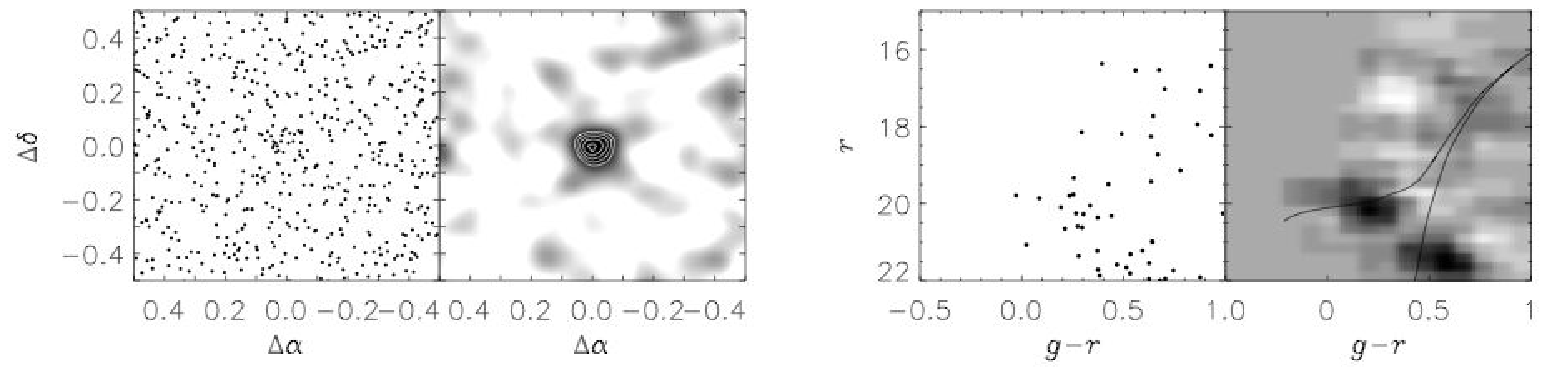} 
 \includegraphics{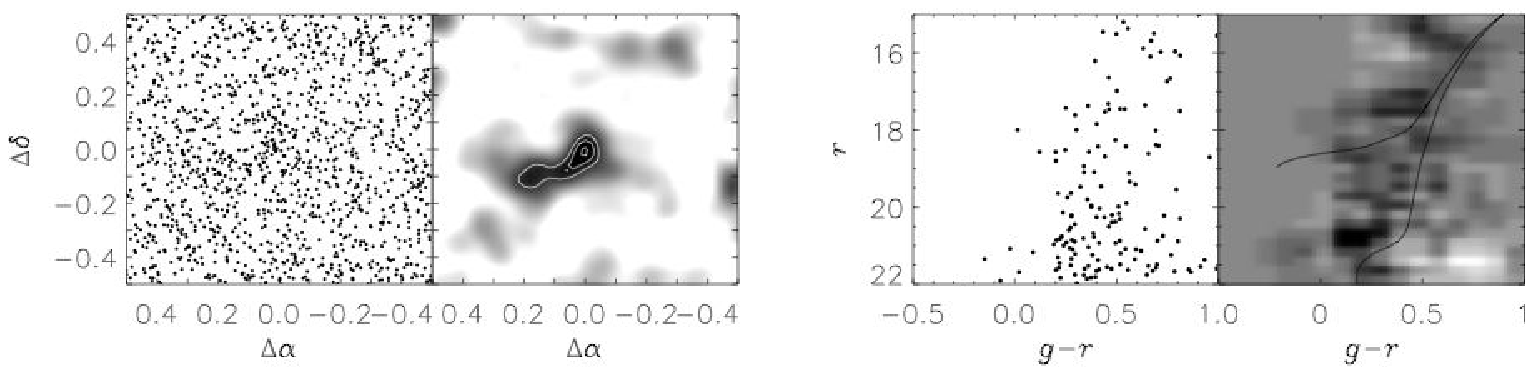} 
\caption{The four unknown detections with detection strength equal to or greater than known satellites (except Leo V). From top to bottom: ``Canes Venatici W'', ``Hercules X'', ``Ursa Major Y'' and ``Virgo Z''. Isochrones show the distance interval at which these overdensities produced the strongest detections.} 
\label{fig:dets5} 
\end{figure*}

Figure \ref{fig:dets5} shows four unidentified overdensities
that have CMDs qualitatively similar to that of a dSph. All show statistically 
significant spatial clustering and do not coincide with a visible overdensity of background galaxies. While several unknown detections have CMDs broadly consistent with old stellar populations, we present here four detections that are as strong or stronger than the detections of UMa and PegDIG. We present their positions and distances estimated from the CMD in Table \ref{tbl:cans}.

\begin{deluxetable}{lrrrr} 
\tablecaption{Positions of Strongest MW Satellite Candidates \label{tbl:cans}} \tablewidth{0pt} \tablehead{ 
\colhead{Designation} & \colhead{$\alpha$} & \colhead{$\delta$}  & \colhead{$(\alpha,\delta$)}& \colhead{$D$ (kpc)}} 
\startdata 
``CVn W''	& 13:16:04.8 & +33:15:00 & $(199.02,33.25)$ & $\sim 160$ \\ 
``Her X''	& 16:27:45.6 & +29:27:00 & $(246.94,29.45)$ & $\sim 100$ \\ 
``UMa Y''	& 12:11:57.6 & +53:35:24 & $(182.99,53.59)$ & $\sim 100$ \\ 
``Vir Z''	& 12:20:19.2 & -1:21:00  & $(185.08,-1.35)$ & $\sim 40$ \\ 
\enddata 
\end{deluxetable} 

%%%%%%%%%%%%%%%%%%%%%%%%%%%%%%%%%%%%%%%%%%%%%%%%%%%%%%%%%%%%%%%%%%%%%%%% 
\section{Exploring Detection Efficiency With Synthetic Satellites}\label{limits} 
The most advantageous aspect of a large, uniform search for MW dwarfs
is the ability to rigorously calculate its detection limits in order
to compare observations with predictions. To calculate the detection
completeness of our search, artificially generated galaxies are
embedded in simulated stellar foreground fields and put through the
detection algorithm to investigate the sensitivity as a function of
galaxy distance, luminosity, scale-length, and Galactic latitude.  In
this section, we describe in detail the method used to synthesize artificial SDSS fields and dSph satellites.

\subsection{Sowing the Simulated Fields}\label{fields}
When simulating fields in which to embed artificial galaxies, our
  goal is to create a large number of fields with the same point
  source color, magnitude, and density distributions as observed in the
  SDSS DR6 footprint.  The detectability of a dSph may change
  depending on its position in the sky. For example, those at low
  Galactic latitudes will be harder to detect than those at high
  latitudes, owing to the greater number of foreground stars. The
  relative proportions of the thin disk, thick disk, and stellar halo
  will also vary with latitude and longitude, changing what fraction
  of foreground stars will be included in the color-magnitude
  selection described in \S3.

\begin{figure}%[!ht] 
 \includegraphics{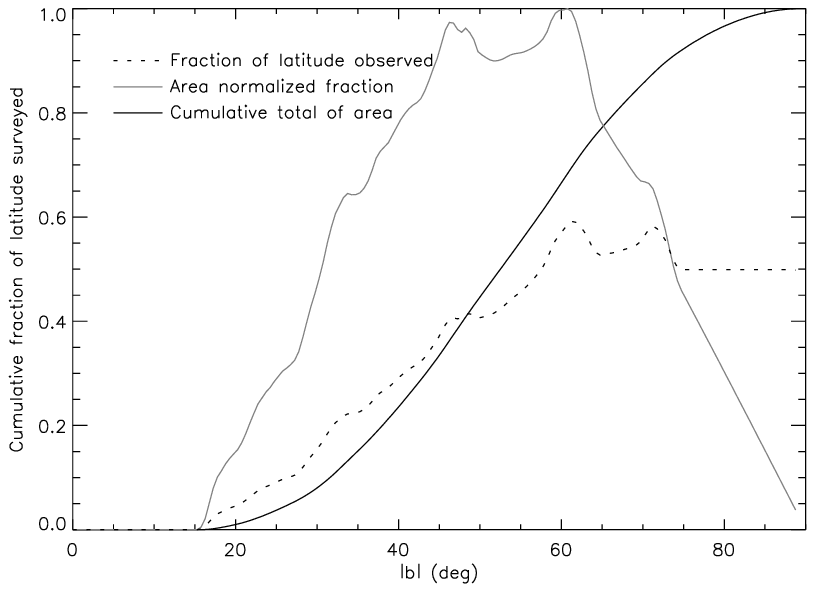} 
 \caption{The fraction of sky as observed by SDSS DR6 at each
   latitude. The dotted line shows what fraction of the small circle
   on the celestial sphere traced by each latitude has been surveyed,
   and the grey line is this fraction weighted by the cosine of
   latitude, to give a relative sky area observed at each
   latitude. The largest area of DR6 observations occur at
   $b\simeq60^{\circ}$. The solid black line is the cumulative total
   of the grey line.} \label{fig:latfrac}
\end{figure}

 To conduct a controlled experiment to see how Galactic foreground
  affects detection efficiency over the DR6 footprint, we first select
  three fiducial latitudes to simulate: the median latitude of the
  DR6 footprint, and the latitudes above and below which 10\% of the
  survey lies. Figure \ref{fig:latfrac} shows the fraction of sky
  observed by SDSS DR6 as a function of latitude (dashed
  line). Weighting this fraction by $\cos b$ gives the relative area
  on the celestial sphere that each observed latitude occupies (solid
  grey line), showing that the majority of the DR6 footprint by area
  is located between $b\approx45^{\circ}$ and
  $b\approx65^{\circ}$. The cumulative total (solid black line) allows
  us to choose the latitudes corresponding to 10, 50 and 90\% levels
  of DR6, namely $31^{\circ}$, $53^{\circ}$ and $74^{\circ}$. These
  are the three values of latitude that we implement in our
  simulations.

\begin{figure}%[!ht] 
 \includegraphics{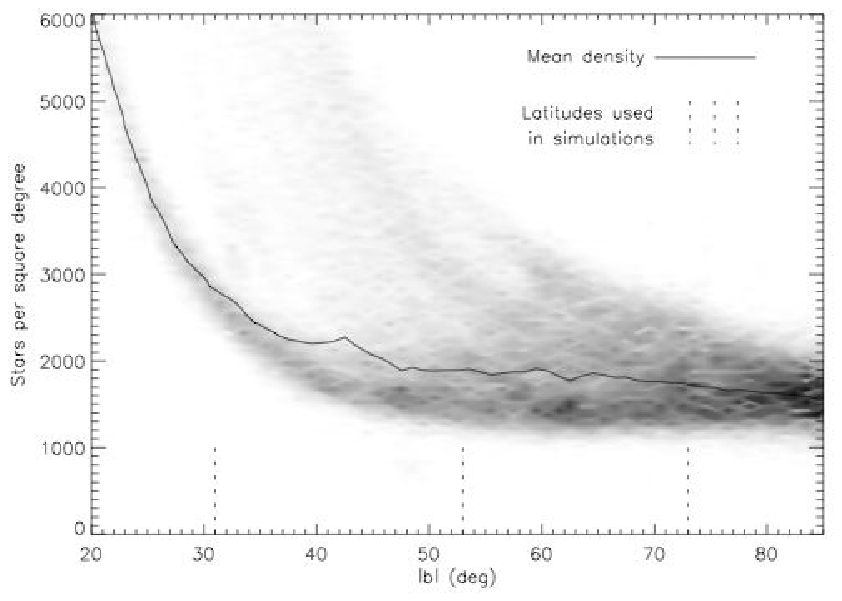} 
 \caption{Grey-scale plot of the number of sources per square degree
   satisfying $r<22.0$ and $g-r<1.0$ versus absolute Galactic
   latitude. The black line traces the mean density at each
   latitude. The three latitudes we choose to simulate are marked
   along the x-axis.} \label{fig:latdens}
\end{figure}

Now that we have chosen what latitudes to simulate, we need to relate
these to the stellar foreground density.  Figure \ref{fig:latdens}
presents a 2D histogram of latitude and foreground density,
considering only stars brighter than $r=22.0$ and bluer than
$g-r=1.0$.  This figure shows a span in foreground levels at each
latitude. The solid black line traces the median and our chosen
latitudes are marked along the x-axis. For each of our latitudes, we
take a slice through the 2D histogram and use this distribution of
densities to randomly assign a density for each of our simulated
fields. Each artificial star in our simulated fields is assigned
photometric parameters from a star in DR6, chosen at random from all
stars within $\pm0.5^{\circ}$ of the latitude in question. These stars
are then randomly distributed in a $3^{\circ}\times 3^{\circ}$ field.

\subsection{Forging Virtual Dwarfs}\label{sims} 
\begin{figure}[!ht] 
 \includegraphics{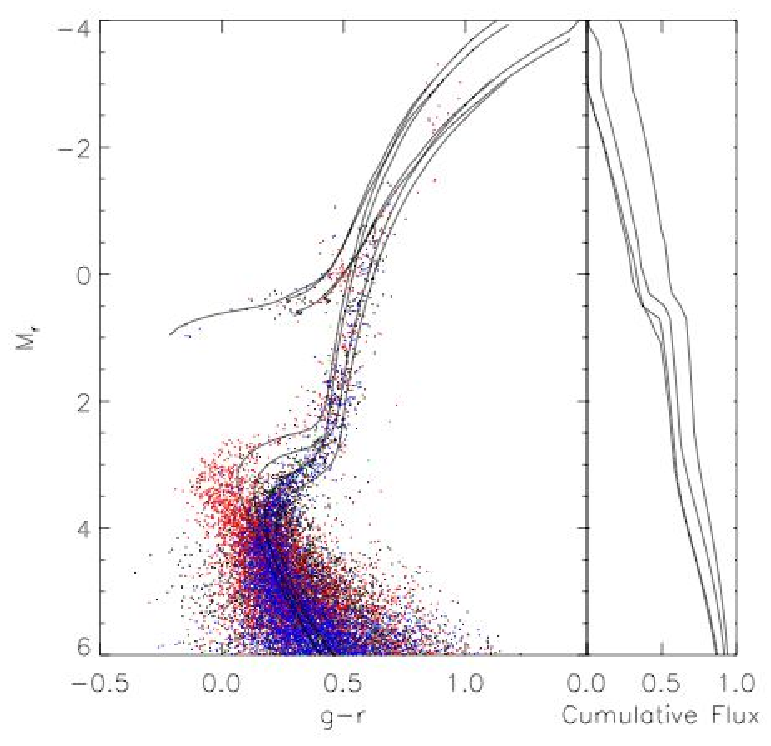} 
 \caption{HST data of three Milky Way satellite dSphs (Carina, Draco, and Ursa Minor; \citealt{hstphot}) with SDSS isochrones
   \citep{sdssiso} overlaid. The right panel shows the cumulative
   luminosity functions for the corresponding isochrones, using the
   four combinations of [Fe/H]$=-2.27,-1.5$ and age$=8,14$Gyr. Data
   are corrected for distance and presented in absolute magnitude. }
\label{fig:hst} 
\end{figure}

To simulate a dSph galaxy CMD, we enlist HST observations of three MW
satellites\footnote{http://astronomy.nmsu.edu/holtz/archival/html/lg.html}: Carina, Draco, and Ursa Minor
\citep{hstphot}. Figure \ref{fig:hst} shows the combined CMD of these objects. We take this
$M_V$ and $V-I$ CMD and translate it in color and magnitude loosely to match the \cite{sdssiso}
isochrones $g$ and $r$. The reasonable agreement between the HST data
after approximate transformation and \cite{sdssiso} isochrones in $g$
and $r$ is sufficient to use the HST data for our simulated
objects.

We use these data to create a composite old stellar
population catalog of stars brighter than $M_r=6$ by combining sources from the three HST dwarfs. Carina, Draco, and Ursa Minor each contribute 5,548, 4,487, and 3,296 stars respectively. Each
time we simulate a dwarf galaxy of $n$ stars, we select those $n$
stars at random from this composite catalog. The luminosity is calculated
from the integrated flux from all stars, and a correction added to
account for stars below an absolute magnitude of $M_{r}=6$. The
cumulative luminosity functions in the right panel of Figure
\ref{fig:hst} show that typically $\sim10\%$ of the total flux
originates from stars below this cut-off. We then adjust the photometry of
the stars to the correct distance modulus and add photometric
scatter to reflect increasing measurement uncertainty with fainter
magnitudes. We do this by finding the best fit for the $1\sigma$
magnitude uncertainty as a function of magnitude for each band in the
SDSS data, and adding a normally distributed random realization
$\delta$ of this value $\sigma(m)$ for the adjusted magnitude, $m_{\rm
  star}=m+\delta\sigma(m)$. We then assign random positions based on a
Plummer surface brightness profile with a specified physical scale
length at a given heliocentric distance. 

\begin{figure}[!ht] 
\center 
 \includegraphics{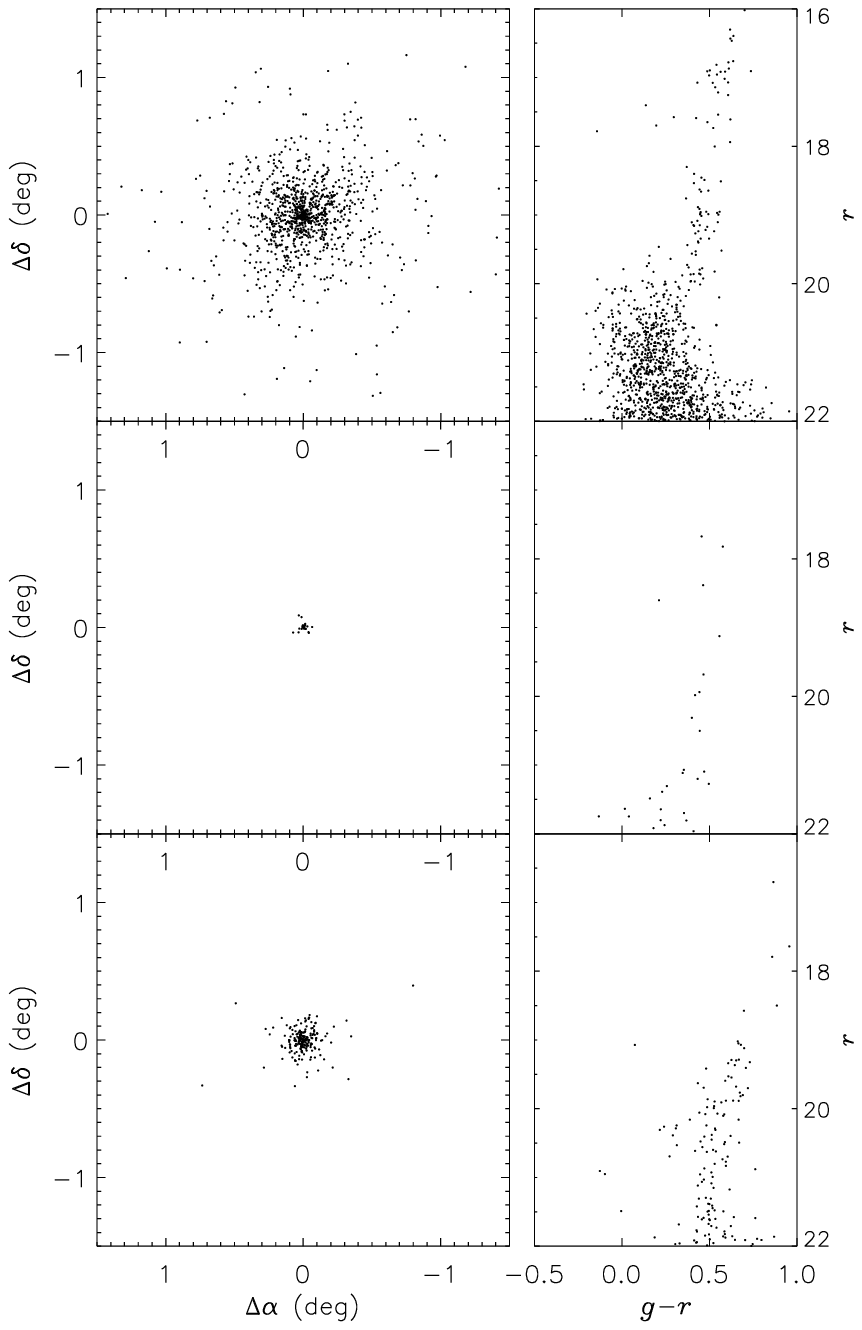} 
\caption{Simulated dSph systems. \\
\emph{Top}: $d=25$ kpc, $r_h=250$ pc, $M_V=-4.7$. \\
\emph{Middle}: $d=45$ kpc, $r_h=40$ pc, $M_V=-2.3$.  \\
\emph{Bottom}: $d=100$ kpc, $r_h=250$ pc, $M_V=-5.6$. }
\label{fig:genplot}
\end{figure}

Figure \ref{fig:genplot} shows examples of three simulated dSphs. The
middle panel shows a system not unlike Bo\"otes II, highlighting the
paucity of stars in the objects we are searching for.  It is important
to note that at these low luminosities, the total luminosity of
galaxies with the same number of stars can vary dramatically, with
this variation increasing for galaxies with fewer stars. A single RGB
star can have a magnitude of $M_{r}=-3$ which is well in the regime of
the total magnitudes of recently discovered satellites.  Each
generated galaxy is embedded in a simulated field, and then processed
as described in \S3.2, \S3.3 and \S3.4.

\subsection{Charting Detection Efficiency}\label{de} 
To test the efficiency of our search algorithm as a function of galaxy
luminosity, scale-length, distance, and Galactic latitude, we generate
a total of 3,825,000 galaxies spanning ranges in Galactic latitude,
luminosity, physical size, and distance. We simulate systems at
latitudes of 31$^{\circ}$, 53$^{\circ}$ and 73$^{\circ}$ and with
$2^x\times100$ stars brighter than $M_V=6$, where $x$ is an integer
between 0 and 11 (giving a range of 100 to 204,800 stars). These
stellar totals correspond to mean total magnitudes of $M_V=-1.5$, $-2.3$,
$-3.1$, $-3.9$, $-4.7$, $-5.5$, $-6.2$, $-7.0$, $-7.7$, $-8.5$, $-9.2$
and $-10.0$. For each of the 36 combinations of latitude and
magnitude, we simulate a large number of galaxies with distances and
physical scale-lengths randomly generated with the limits
$1.3<\log{d/kpc}<3.0$ and $0.9<\log{r_h/pc}<3.0$. For the brightest
and faintest systems we tailor these ranges to avoid redundant
iterations; there is little to be gained by simulating a $M_V=-1.5$
system at 200 kpc or a $M_V=-10.0$ system at 20 kpc. Hence, the total
number of simulations for each magnitude/latitude combination varies,
but is chosen such that there are typically 500 simulations in each
0.1 log($d$) $\times$ 0.1 log($r_h$) bin of Figure
\ref{fig:detfrac}.

%%%%%%%%%%%%%%%%%%%%%%%%%%%%%%%%%%%%%%%%%%%%%%%%%%%%%%%%%%%%%%%%%%%%%%%%%%%%%%%%%%%%%%%%%%%%%%%
\section{Dissecting Efficiency Trends}
\begin{figure*}%[!ht] 
 \center 
 \includegraphics{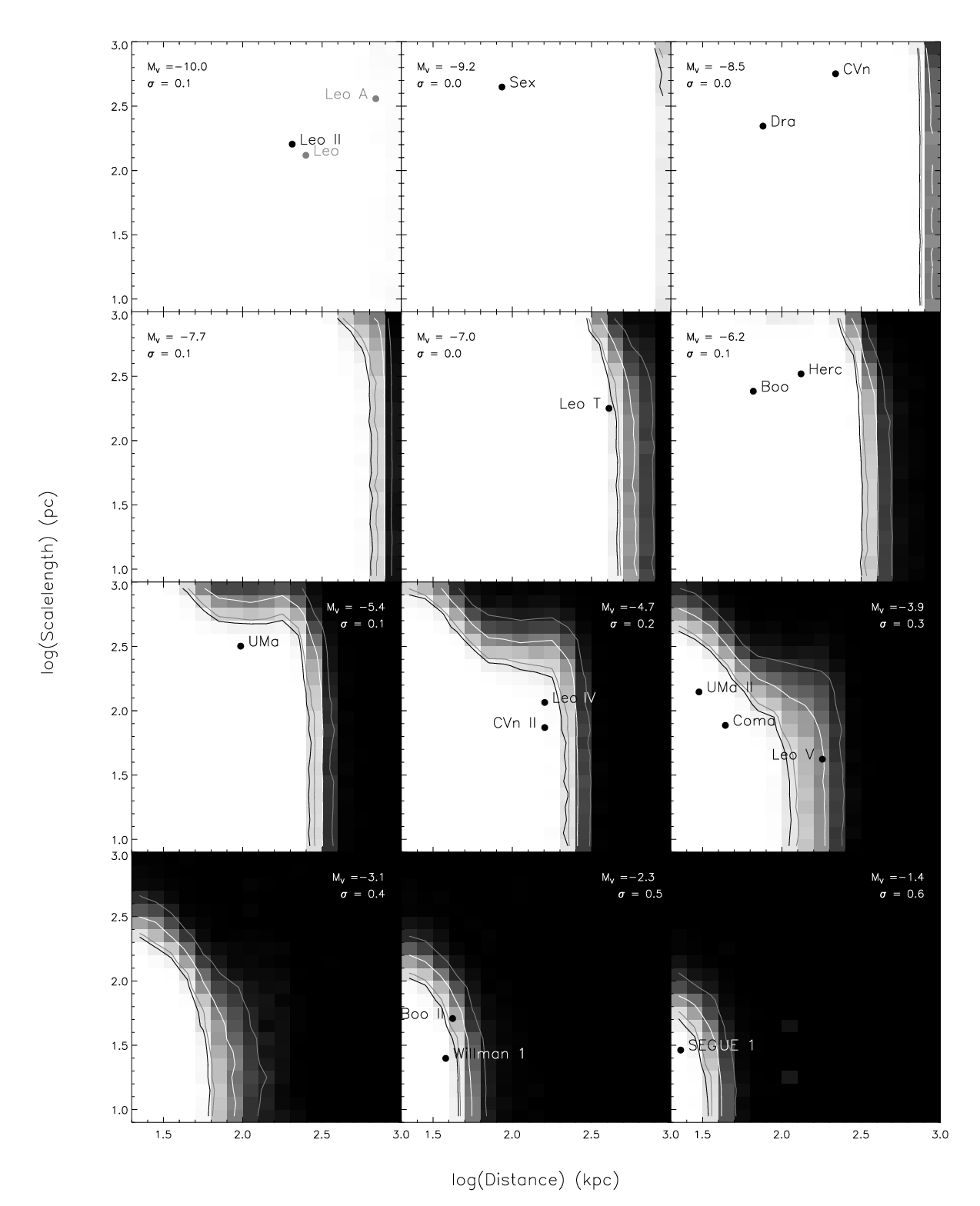} 
 \caption{Detection Efficiency for specific galaxy parameters. Each 
 panel shows the detection efficiency as functions of distance and 
 scale-length for a particular number of galaxy stars. The average 
 total absolute magnitude for each of these sets of galaxies is shown 
 along with the standard deviation in magnitudes. Contours show the 90\%, 84.13\%, 50\%, and 15.86\% levels. Sizes and distances of known MW dSphs are shown in the best matching magnitude panel. Leo and Leo A are shown in grey as they are significantly brighter than $M_V=-10.0$. Values for newly discovered objects plus Draco are taken from \cite{martinhood} except for Leo V \citep{leov}. Values for Leo, Leo II, Leo A, and Sextans taken from \cite{mateo}.} 
 \label{fig:detfrac} 
\end{figure*}

Figure \ref{fig:detfrac} shows the detection efficiency of simulated
dwarf galaxies as a function of luminosity, scale-length, and distance
at the median SDSS Galactic latitude ($53^{\circ}$).  Overplotted are
each of the Milky Way satellites detected in SDSS, not including
Koposov 1 and 2. Each panel contains a grey-scale map of the
detection efficiency for simulated galaxies of the mean absolute
magnitude specified in the panel. Because the total magnitude varies
for systems with a constant number of stars, we quote both the mean
magnitude and the standard deviation of magnitudes for each
panel. White shows regions of 100\% efficiency, while black shows
0\%. The four contours, moving outwards from 100\% efficiency, show
the 90\%, 84.13\%, 50\%, and 15.86\% levels. The 84.13\% and 15.86\%
levels were chosen to illustrate the $\pm1\sigma$ in detectability as
a function of distance and size.

The greater the number of stars in a simulated galaxy, the less its
absolute magnitude will vary between realizations, so the standard
deviation in the integrated magnitudes of simulated galaxies
contributing to each panel decreases with increasing magnitude. The
$M_V=-7.7$ panel which shows a small increase in standard deviation
marks a change in the way the galaxies are simulated; we are now
simulating systems with more stars than are in our HST catalog so the
simulated stars no longer have unique photometry drawn from this
catalog. This amplifies the small number effect of single stars on the
total magnitude.

Figure \ref{fig:detfrac} illustrates the necessity for a large
  number of simulations as there are subtle features that would
  otherwise be unresolved. For example, this figure shows that the
  detectability of dwarfs is not a step function in distance, but
  rather slowly falls off at a rate that differs for systems with
  different total luminosities. This is in contrast to \cite{koplf}, who found a steep boundary between 100\% and 0\% efficiency. The gradual fall-off in dwarf detectability with distance will be discussed in more detail
  in \S\ref{dist}, but we postpone detailed comparison with \cite{koplf} until \S\ref{comparison}.

The critical factor affecting the detectability of an object with our algorithm is the
number of stars brighter than $r=22.0$ that fall under the $r_h=4.5'$
Plummer smoothing kernel. In the following sections, we use this to gain physical understanding of the features of Figure \ref{fig:detfrac} and to derive an analytic expression to describe detection efficiency as a function of galaxy magnitude, size, distance, and Galactic latitude. This analytic expression, as well as a routine to interpolate detectability directly from the simulations, will be made publicly available and can be used, for example, to correct the MW satellite luminosity function as previously done by \cite{koplf} or make an estimate on the corrected radial distribution of MW satellites. Such endeavours are beyond the scope of this paper, but we use the function to estimate the total number of MW satellites that remain undetected, presented in \S\ref{miss}. 

\subsection{Efficiency versus Distance}\label{dist}
Figure \ref{fig:detfrac} shows that the detectability of resolved
  dwarfs around the Milky Way is not a step function in distance. As
  distance to a dwarf galaxy increases, the number of stars brighter
  than $r=22.0$ ($N_{r<22}$) decreases. In an idealized scenario, the
  detectability of that dwarf would drop from 100\% to 0\% at a
  distance beyond which the number of resolved stars required to
  produce a detection is larger than $N_{r<22}$.  As we have discussed
  in previous sections, random variations in the stellar luminosity
  function can be substantial in the faintest systems, hence
  $N_{r<22}$ will be affected by stochastic fluctuations. Moreover,
  the wide range in foreground densities at a given Galactic latitude
  (see Figure \ref{fig:latdens}) impacts the detectability of two identical dwarfs.
  Therefore the transition from 1.0 to 0.0 detection efficiency is not
  expected to be a step function, but rather described by a Gaussian integral. \cite{koplf} also modeled the detectability transition with a Gaussian integral, despite finding a steep decline. So
  detection efficiency as a function of log(distance), which for brevity we denote $_Ld$ can be described
  as
\begin{eqnarray*}
DE(_Ld) = erf\left(\frac{_L\bar{d}-_Ld}{\sigma_{_Ld}}\right) \nonumber
\end{eqnarray*}
where $_Ld$ is the logarithm of distance, and $_L\bar{d}$ and $\sigma_{_Ld}$ are the mean and standard deviation of log(distance). The mean corresponds to the distance at which a system would be detected with 50\% efficiency. The error function $erf$ is
defined as:
\begin{eqnarray*}
erf(x) = \frac{2}{\sqrt(\pi)}\int^{x}_{0}e^{-t^2}dt. \nonumber
\label{eqn:erf}
\end{eqnarray*}

Examination of Figure \ref{fig:detfrac} shows that $\sigma_{_Ld}$ (how
quickly efficiency transitions from unity to zero) changes depending
on luminosity. We would naively expect $\sigma_{_Ld}$ to continue
increasing with decreasing brightness as small number statistics
becomes more dominant. Instead it shows a maximum at $M_V\approx-3.9$
before decreasing. This is a result of the stochastic fluctuations and
the derivative of the luminosity function; since $N_{r<22}$ varies for
systems with the same total number of stars, the individual distance
that each of these systems could be detected at also varies. As the
number of stars above the brightness limit is dependent on the LF, the
slope of the LF determines how $N_{r<22}$ changes with distance. Hence $\sigma_{_Ld}$ is smaller for the faintest objects when the MSTO is required for a detection because the LF is at its steepest at the turnoff.

\subsection{Efficiency versus Scale-length} 
The fraction of a
  dwarf's stars within our $4.5'$ spatial smoothing kernel decreases
  with increasing physical scale-length and/or decreasing distance.
A system of some luminosity and distance that is detectable when its
angular size is $\lesssim 4.5'$ may thus be undetectable if those same
stars are spread over a larger angular scale. As the concentration of stars increases we would expect detection efficiency to also increase. However, when the angular size of a dwarf is comparable to the smoothing kernel size,
the detectability does not appreciably improve with further decrease
in size since the number of stars within the Kernel is not significantly changing. Hence objects of this angular size or smaller will be
detected with the same efficiency. 

Once the angular size becomes
larger than the kernel size, the number of stars within the kernel declines. The
relationship between size and detection distance is dependent on the
stellar luminosity function of the system. Take for example an object
with an angular size larger than the smoothing kernel, detected with
50\% efficiency at some distance. To keep the object at 50\% efficiency as we continue to increase the physical size,
the drop in efficiency can be counteracted by decreasing the object's
distance. As this object is moved closer the number of stars above
$r=22.0$ increases at a rate corresponding to the LF. At
the distance when the main sequence turn-off (MSTO) becomes brighter than
$r=22.0$ ($\sim65$ kpc), the rapid increase in the number of stars
corresponds to a sudden improvement in detection efficiency, evident
in Figure \ref{fig:detfrac} at $\log(d)\approx1.8$ in the $M_V=-5.5$, $-4.7$ and $-3.9$ panels. As with distance, efficiency versus scale-length can be modeled by a Gaussian integral,
but with the mean and standard deviations as functions of distance;
so
\begin{eqnarray*}
DE(_L r_h) = erf\left(\frac{_L\bar{r}_h(_L\bar{d})-_Lr_h}{\sigma_{_Lr_h}(_Ld) }\right). \nonumber
\end{eqnarray*}

\subsection{Analytically Expressing Detection Efficiency} 
\begin{figure*}%[!ht] 
 \center 
 \includegraphics{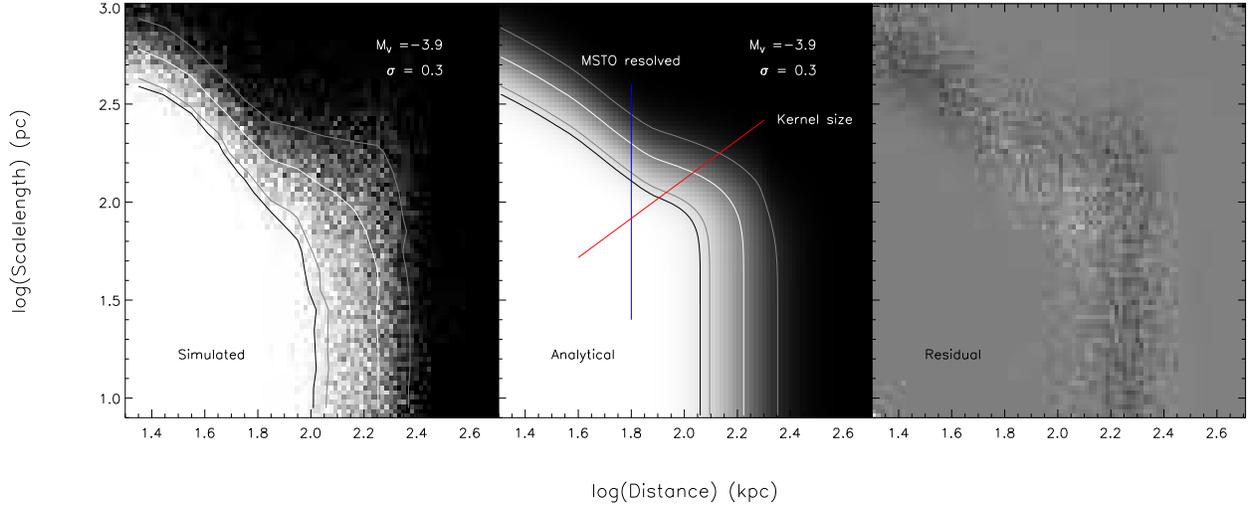} 
 \caption{\emph{Left}: $M_V=-3.9$ panel of Figure \ref{fig:detfrac} binned to higher resolution. \emph{Center}: Same as left panel but using the analytical expression to estimate detection efficiency. \emph{Right}: Residual of the model-analytical efficiency.} 
 \label{fig:detfracsim} 
\end{figure*}

Combining the previous two results, we can analytically describe detectability with a
Gaussian integral over $\log{d}$ multiplied by another Gaussian
integral over $\log{r_h}$. To introduce magnitude $M_V$ and latitude
$b$, we set the means ($_L\bar{d}$, $_L\bar{r}_h$) and standard deviations
($\sigma_{_Ld}$, $\sigma_{_Lr_h}$) in the integrals to be functions of
$M_V$ and $b$. Therefore the detection efficiency $DE$ can be
expressed:
\begin{eqnarray*}
DE = erf\left(\frac{_L\bar{d}(M_V,b)-_Ld}{\sigma_{_Ld} (M_V,b)}\right)\times erf\left(\frac{_L\bar{r}_h(_L\bar{d}(M_V,b))-_Lr_h}{\sigma_{_Lr_h}(_Ld(M_V,b))}\right) \nonumber
\label{eqn:de}
\end{eqnarray*}
The means and standard deviations can be found by fitting Gaussian integrals along the distance and scale-length axes of the panels in Figure \ref{fig:detfrac}. The main source of uncertainty in this expression is the galaxy
luminosity, which while correlated with the number of stars, can vary
by over a magnitude for systems of equal detectability. The function
does however give a good statistical approximation from which to
estimate the properties of the true Milky Way satellite
population. In Figure \ref{fig:detfracsim} we compare the $M_V=-3.9$ panel of Figure
\ref{fig:detfrac} with the analytical function. There is good
agreement between the empirical and analytical detection efficiencies with a $1\sigma$ deviation of only $\sim8.7$\% across the
entire range of parameters. For reference the size of the $4.5'$ smoothing kernel and the distance at which the MSTO becomes resolved are shown in the center panel in red and blue respectively. 

The analytical efficiency is compared in
Table \ref{tbl:dets} to the interpolated efficiency from the grids in
Figure \ref{fig:detfrac} for real MW dwarfs. For all objects besides
Boo II, Leo V, and Leo T, the difference is within $\sim1$\%. At first
inspection it may seem odd that all objects have a very high, almost
100\%, efficiency, but given that most of the parameter space probed
by our simulations yields either zero or unity efficiency, it is not
unexpected that the handful of objects detected in this vast volume
are detected with high efficiency.
 
\begin{deluxetable}{lrrr} 
\tablecaption{Comparison of Interpolated and Analytical Detection Efficiencies \label{tbl:dets}} \tablewidth{0pt} \tablehead{ 
\colhead{Object} & \colhead{Interp.} & \colhead{Analyt.} & \colhead{Diff.}} 
\startdata 
Boo	& 99.84 & 99.98 & $-0.14$ \\ 
Boo II	& 90.39 & 95.96 & $-5.57$ \\ 
CVn	& 100.0 & 100.0 & $0.0$ \\ 
CVn II	& 98.26 & 99.07 & $-0.81$ \\ 
Com	& 99.71 & 100.0 & $-0.29$ \\ 
Dra	& 100.0 & 100.0 & $0.0$ \\ 
Her	& 99.82 & 99.83 & $-0.01$ \\ 
Leo	& 100.0 & 100.0 & $0.0$ \\ 
Leo II	& 100.0 & 100.0 & $0.0$ \\ 
Leo A	& 100.0 & 100.0 & $0.0$ \\ 
Leo IV	& 98.65 & 99.77 & $-1.12$ \\
Leo V	& 83.56 & 91.27 & $-7.71$ \\  
Leo T	& 93.41 & 99.38 & $-5.97$ \\ 
Segue 1	& 100.0 & 100.0 & $0.0$ \\ 
Sex	& 100.0 & 100.0 & $0.0$ \\ 
UMa	& 99.97 & 99.86 & $0.11$ \\ 
UMa II	& 100.0 & 99.96 & $0.04$ \\ 
Will 1	& 98.56 & 99.30 & $-0.74$ \\
 \enddata 
\end{deluxetable} 

\subsection{Efficiency versus Latitude}\label{latitude}
\begin{figure}%[!ht]
 \center
 \includegraphics{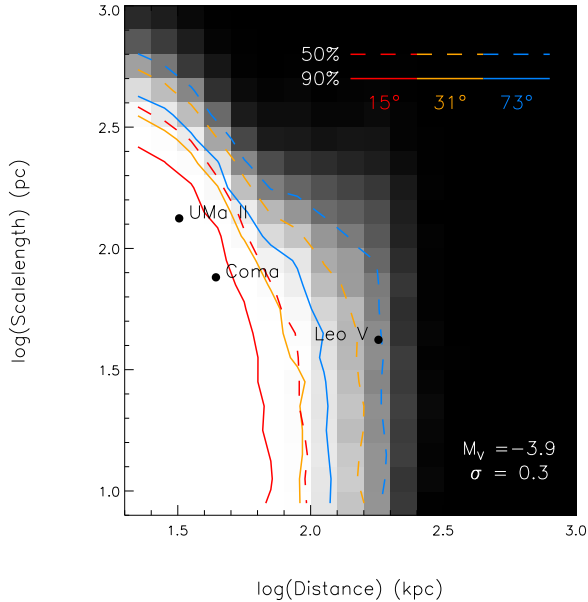}
 \caption{The same as the $M_V=-3.9$ panel of Figure
   \ref{fig:detfrac}, but showing the 50\% and 90\% contours of the
   simulations at $b=74^{\circ}$ (blue), $31^{\circ}$ (orange), and
   $\sim15^{\circ}$ (red).}
 \label{fig:detboth}
\end{figure}

Unlike the other parameters which vary a dSph's signal strength,
Galactic latitude affects detection efficiency by changing the
foreground density, and therefore noise above which we must detect a
signal. If latitude plays a significant role in the detectability of
dwarfs, then it must be taken into account when making any corrections
to the MW satellite census. Figure \ref{fig:detboth} shows the
$M_V\approx-3.9$ panel of Figure \ref{fig:detfrac} ($b=53^{\circ}$)
with the addition of the 50\% detection efficiency (dashed) and 90\%
detection efficiency (dotted) contours of the $b=31^{\circ}$ (orange)
and $74^{\circ}$ (blue) simulations overplotted. As expected, an
object of given size and luminosity will not be detectable as far away
at low latitudes as it would be at closer to the Galactic pole. For
example, an object with $r_h\approx30$ pc and $M_V\approx-3.9$ at
$b=74^{\circ}$ can be detected with 90\% efficiency as far as
$\sim120$ kpc, while the same object at $b=31^{\circ}$ has a 90\%
efficiency at $\sim95$ kpc.

To anticipate the effect that varying Galactic foreground will
  affect future dwarf searches in data that goes closer to the
  Galactic plane than SDSS, we also repeat the simulation of
  $M_V\approx-3.9$ galaxies at a foreground density of 10,000 stars
  per square degree, approximating a latitude of $\sim15^{\circ}$.
  The 50\% and 90\% detection efficiency contours (red) for these
  simulations are also shown on Figure \ref{fig:detboth}. This
further reduces the 90\% detection distance of our example object to
$\sim80$ kpc but demonstrates that future surveys should still detect
dwarfs at relatively low Galactic latitudes, barring extinction
effects.

 \begin{figure}%[!ht]
 \center
 \includegraphics{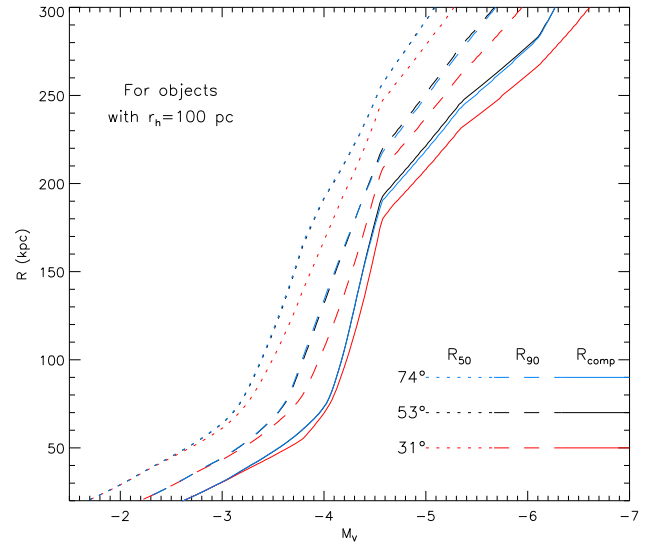}
 \caption{50\%, 90\%, and 99\% completeness distances for a $r_h=100$
 pc object as a function of magnitude at three Galactic latitudes:
 $31^{\circ}$, $53^{\circ}$ and $71^{\circ}$. The $53^{\circ}$ and
 $71^{\circ}$ curves are virtually indistinguishable showing that
 latitude does not significantly impact satellite detection over
 latitude ranges of DR6. }
 \label{fig:rmax5}
\end{figure}

To check that latitude has the same lack of effect over different
magnitudes, we calculate the 50\%, 90\%, and 99\% detection efficiency
distance for a $r_h=100$ pc object over the magnitude range
$-1.5<M_V<-7.0$ at $b=31^{\circ}$, $53^{\circ}$, and $71^{\circ}$
(Figure \ref{fig:rmax5}, note that in this figure the distance scale
is linear). The $53^{\circ}$ and $71^{\circ}$ curves are
indistinguishable and the $b=31^{\circ}$ curve is typically less than
20 kpc lower. This demonstrates that over the DR6 footprint, latitude
does not play an important role on average in the detectability of
objects.  However, if we are unlucky, then individual objects could by
chance lie in directions of unusually high foreground counts.

\subsection{Comparing $R_{99}$, $R_{90}$ and $R_{50}$}
\begin{figure}%[!ht]
 \center
 \includegraphics{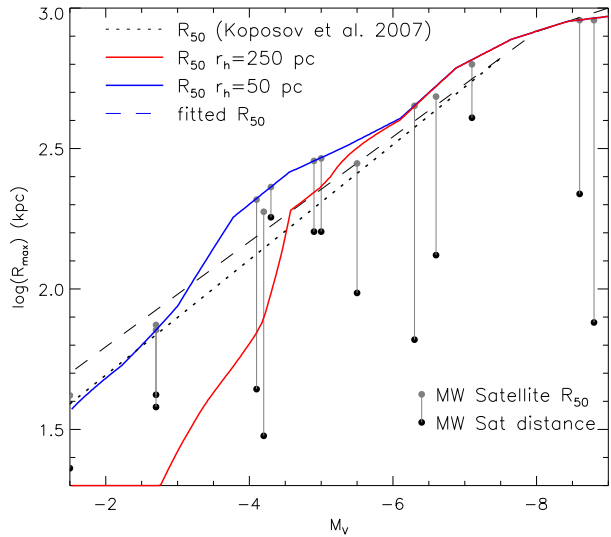}
 \caption{Comparison of the 50\% detection distance as a function of
 magnitude for K07 (dotted) and for our analytical efficiency using
 $r_h=250$ pc (red) and $r_h=50$ pc (blue). MW dwarfs are shown as
 filled circles. }
 \label{fig:rmax}
\end{figure}

We use the analytical expression derived in \S6.4 to estimate the
distance at which each of the Milky Way dwarfs would be detected with
50\% efficiency, $R_{50}$ (Figure \ref{fig:rmax}, grey dots). Because
$R_{50}$ depends on half-light radius, as well as luminosity, we also
for reference show $R_{50}$ for objects with $r_h=250$ pc (red) and
$r_h=50$ pc (blue).  These lines show that the detectability of the
lowest luminosity dwarfs is severely reduced for large scale sizes.
Objects with Segue 1, Bo\"otes II, or Willman 1-like luminosities
would not have been detectable with scale sizes of 100 pc or larger,
even at very nearby distances.  This size bias is important to bear in
mind, particularly given that the three M31 satellites discovered by
\cite{lgtrio} that highlight regions of dwarf galaxy parameter space that
have not previously been observationally seen.

Assuming the size-luminosity distribution of known satellites is
representative of all satellites we can ignore size and approximate
$R_{50}$ as a function of $M_V$ by linear fit to the $R_{50}$ of
actual dwarfs:
\begin{eqnarray*}
\log R_{50} = -0.187 M_V + 1.420.
\end{eqnarray*}
For comparison, the \cite{koplf} equivalent of $R_{50}$ is shown
in Figure \ref{fig:rmax} (dotted line) obtained from Table 3 in their paper, and discussed further in \S
\ref{comparison}. For reference, the actual distances to the MW dwarfs are
shown as black dots.

A more useful quantity might be $R_{complete}$, the maximum distance
at which objects can be detected. Although we choose to use
``complete'' to refer to 90\% detectability, complete could be defined
as, say 90\% or 99\% efficiency. This can also be approximated by a linear fit to the results for the actual MW dwarfs:
\begin{eqnarray*}
\log R_{90} = -0.204 M_V + 1.164
\end{eqnarray*}
or
\begin{eqnarray*}
\log R_{99} = -0.217 M_V + 1.005.
\end{eqnarray*}

These relationships again assume that the known satellites are typical
of the MW satellite population as a whole, i.e. objects with
luminosities comparable to Segue 1 are similar in size to Segue 1 and
not significantly larger.

Returning to Figure \ref{fig:rmax5} we see the distance range over
which detectability changes from 99\% to 50\%. $R_{90}$ is typically
$\sim20$ kpc closer than $R_{50}$, and $R_{99}$ is $\sim20$ kpc closer
still. From this figure we can also see that objects brighter
than $M_V\approx-6.5$ mag ($M_V\approx-5.9$ mag) are detected with 99\% (90\%)
efficiency out to 300 kpc. We can infer from this that all dwarfs
within the MW virial radius brighter than $M_V\approx-6.5$ mag are
known, and any satellites still undetected are likely to be
reminiscent of objects like Coma Berenices, Bo\"otes II or Segue 1 at
distances greater than $\sim40$ kpc. An ultra-faint satellite such as
Segue 1 can only be detected with 50\% efficiency out to $\sim40$ kpc;
there may be many more such objects beyond this distance.

The code for interpolating detection efficiency from our simulations
as well as the analytical function will be made available for download
at the Astronomical Journal website. Interpolation will give more
accurate results, but the analytical function will provide flexibility
for customization to suit individual needs and implementation of any
future improvements.

\subsection{Caveats}
An underlying assumption of our simulations is that the DR6 point
source catalog is uniformly 100\% complete to $r=22.0$. This
assumption may result in optimistic detection efficiency estimates for
the faintest and furthest systems. These faintest and more distant
systems would also be subject to the human element; a real object may
detected by the algorithm but on visual inspection be disregarded as
background galaxy cluster or other contaminant. Finally, sources in
our simulated dSphs are distributed circularly
symmetrically. \cite{martinhood} find that the ultra-faint satellites
are in fact quite elliptical, which due to our circular Plummer
smoothing kernel, possibly results in overestimated efficiencies for
objects like Hercules, UMa and UMa II with ellipticities of 0.68, 0.80
and 0.63 respectively. 

%%%%%%%%%%%%%%%%%%%%%%%%%%%%%%%%%%%%%%%%%%%%%%%%%%%%%%%%%%%%%%%%%%%%%%%%%
 \section{Comparison With Koposov et al. (2008)}\label{comparison}
Besides \cite{Willman02} from which this work follows, two other
surveys have recently uniformly searched SDSS for MW satellites,
namely \cite{liu} and \cite{koplf}. \cite{liu} conducted a
straightforward search and presented five satellite
candidates. \cite{koplf} present a study comparable to this work, and
follows from the ``Field of Streams'' \citep{fieldofstreams} that led
to the discoveries of nine of the new Milky Way satellites.  Here we
compare our work in detail with \cite{koplf} and summarize the main
differences in Table \ref{tbl:comp}.
 
The aim of \cite{koplf} was to present a luminosity function of the MW
satellites corrected for luminosity bias. Their analysis discovered
two new extremely faint globular clusters, Koposov 1 and 2
\citep{k12}.  In principle, our analysis is quite similar to
\cite{koplf}, henceforth K07, in that they apply a color cut, smooth
the stellar counts and look for statistically significant
overdensities. There are several distinctions however that we detail
below.

K07 employed a $g-r<1.2$ color cut to remove a substantial fraction of
MW foreground stars and a $r<22.5$ cut to limit the influence of
background galaxies and increasing uncertainties/incompleteness. Our
color-magnitude cuts are tailored to old stellar populations at 16
different distances which serve to eliminate more foreground stars
than the looser K07 cut. The looser K07 color cut leaves enough stars
that a complicated set of detection thresholds is unnecessary, whereas
we must consider the effects of non-Gaussianity in low densities (see
\S\ref{threshold}). K07 deals with background galaxy clusters, a major
source of contaminant overdensities, by producing a galaxy clustering
significance in the same manner as the stellar clustering; anywhere
that a stellar overdensity occurs without a corresponding galaxy
overdensity is much more likely to be a true stellar overdensity. Our
algorithm only includes stars as faint as $r=22.0$ and as such we have
fewer galaxy cluster contaminant detections.

The most substantial difference between our work and K07 is how we
derive the detection limits of our algorithms. Like our work, K07
simulated artificial galaxies to explore the detection efficiency as a
function of size, distance and luminosity. K07 simulated 8,000
galaxies over a similar range of parameters as our study, but with
only $\sim8$ objects per $0.3$ log($d$) $\times$ 0.3 log($r_h$)
$\times$ 0.8 mag bin. There is considerable noise evident in the
detection limits (their Figure 6), and all of the new satellites
appear to lie on the edge of detectability.  K07 observed a steep, but
finite, transition from unity to zero detection efficiency which they
attributed to the large range of distances that fall within each
size-luminosity bin.  However, as discussed in our \S\ref{sims}, a
large number of simulated galaxies are essential for each permutation
of dwarf galaxy parameters to effectively map their detectability.
Our high resolution detection maps ($\sim500$ objects per $0.1$
log($d$) $\times$ 0.1 log($r_h$) $\times$ 0.8 mag bin, $\times3$
latitudes) show that the detectability of a dwarf drops off slowly
with size and distance, and that only Leo T, Leo IV, Leo V, Boo II and
Willman 1 lie close to the edge of detectability. The difference between
90\% and 10\% efficiency typically occurs over 0.2 dex in distance
(kpc) and 0.3 dex in size (pc; see Figure \ref{fig:detfrac}).

Both the K07 and our detection limit calculations suffer from the
implicit assumptions that the SDSS point source catalog is complete to
the photometry limit and that dwarfs are circularly symmetric. These
two assumptions yield detection limits that may be
optimistic. The K07 study includes stars to a limiting magnitude of
$r=22.5$ mag, a half magnitude fainter than our limit of $r=22.0$ mag.
We thus expect that the completeness assumption may impact their
calculated limits more than ours.

To directly compare the effectiveness of both algorithms we return to
Figure \ref{fig:rmax}, showing the distance at which an object is
detected with 50\% efficiency, $R_{50}$ as a function of
magnitude. The dashed line shows the K07 $R_{50}$ which was determined
by fitting a limiting magnitude and surface brightness to the seven
distance panels in their Figure 10. The dots show our $R_{50}$ derived
from the analytical efficiency function for each of the MW dwarfs in
DR6, while the red and blue curves show $R_{50}$ calculated for
objects of $r_h=250$ pc and $r_h=50$ pc respectively. 

Although this comparison roughly shows that we have comparable limits, our calculated detection efficiencies of each dwarf are all greater than 90\% while Table 2 of
K07 lists efficiencies as low as 47\% (neglecting Boo II). While Bo\"otes II is not detected with the standard algorithm of K07,
it is a comparatively strong detection in our algorithm.  We note that the tabulated K07 efficiencies for the known MW dwarfs appear inconsistent with their fitted $R_{50}$, which places some dwarfs much closer than $R_{50}$ than their actual efficiencies would indicate.  The increased dwarf detectability of our survey is owing to a combination of different
techniques and our less stringent detection threshold.  We set our
thresholds to strictly eliminate truly random false positives expected
while still yielding new candidates, and hence have $\sim30$ unknown
detections above our thresholds. Although upon visual inspection of
their CMDs many of these detections appear unlikely to be new dSph
satellites, they may also be tidal debris or distant galaxy
clusters. However, K07 set their detection thresholds just loose
enough to retain all known objects; UMa is their weakest detection and
there are only three unknown detections above this threshold.

\begin{deluxetable*}{p{2.0in}p{2.2in}p{2.2in}}
\tablecaption{Summary of Comparison with Koposov et al. (2008). \label{tbl:comp}} \tablewidth{0pt} \tablehead{ &
 \colhead{This work} & \colhead{Koposov et al. (2007)}}
\startdata
Survey Area & DR6 - 9,500 sq deg & DR5 - 8,000 sq deg \\
Source cuts & Isochrone template at 16
 dist intervals & $g-r<1.2$ and $r<22.5$ \\
Smoothing kernel & $4.5'$ Plummer profile &
 $\sigma=2',4',8'-\sigma=60'$ Gaussians \\ Threshold & Multiple, function of foreground density & Fixed,
 considers background galaxies \\ Modeled detection limits &HST obs of 3 MW dSphs &
 M92 locus \\ Number of Simulations & 3,825,000 & 8,000 for
 general simulation + 1,000 each for known dwarfs within DR5 \\
Efficiency map bin size      (log$(r_h)\times$ log$(d)\times M_V$) & $0.1\times0.1\times0.8$  & $0.3\times0.3\times0.8$ \\
Simulation density     (n per $0.3\times0.3\times0.8$ bin) & $\sim4500$ ($\times3$ latitudes)  & $\sim8$ \\
 \enddata
 \end{deluxetable*}
 
%%%%%%%%%%%%%%%%%%%%%%%%%%%%%%%%%%%%%%%%%%%%%%%%%%%%%%%%%%%%%%%%%%%%%%%%%%%%
\section{The Still Missing Satellites}\label{miss}
A substantial driving force of this work is the Missing Satellite
problem, which the discovery of so many new objects in the space of
three years has shown is far from observationally exhausted. There are
still large regions of parameter space where objects are undetectable,
so there can easily exist more objects within the DR6 coverage that
remain hidden. Future surveys such as the Stromlo Missing Satellites
(SMS) Survey and Pan-STARRS may be able to detect some of these
objects, and we can use our model of detectability to estimate how
many there may be.

We use a simplified version of the approach used in \cite{tollerud}. We first assume that the radial distribution of dwarf
galaxies matches that of all well-resolved subhaloes of the Via Lactea
simulation \citep{vialactea}.  \cite{tollerud} discusses this
assumption in detail; we realize this may not reflect the true MW
dwarf distribution, but our qualitative results are fairly robust to
the assumed profile. For each satellite detected in DR6, we then
determine $R_{90}$ (or $R_{99}$), the maximum distance to which a
satellite of similar properties would be detected with 90\% (99\%)
efficiency. For each value of $R_{90}$ ($R_{99}$), we determine from
the Via Lactea subhalo radial profile what fraction of satellites
should be within this distance, and weight each satellite
accordingly.

Using all dark matter subhaloes with more than 1000 particles within
$r_{vir}=289$ kpc, and adopting a MW virial radius of 258 kpc
\citep{klypin02}, we estimate $\sim13$ ($\sim24$) satellites within
the MW virial radius in the DR6 footprint. Twelve of these would be
the known objects Bo\"otes, Draco, Canes Venatici I and II, Coma
Berenices, Leo I, Leo II, Leo IV, Leo V, Hercules, Ursa Major, and Ursa Major
II, leaving one (12) possible missing satellite(s). From our simulations,
objects brighter than $M_V=-6.5$ mag are detectable with $>99$\%
efficiency out to the virial radius, so we would expect that a
relatively small number of the faintest systems are missing. These
missing satellites may be amongst our candidates, or be Coma
Berenices like or
fainter objects in the outer halo. Whether or not future searches
reveal such objects may validate the assumed radial distribution. If
we assume an isotropic sky distribution of satellites, $\sim13$
($\sim24$) objects within DR6 equates to $\sim52$ ($\sim96$) across
the whole sky.

 If we include the ambiguous objects Segue 1, Willman 1, and Bo\"otes II
 1 in the calculation, then the $R_{90}$ ($R_{99}$) DR6 estimate would be $\sim56$ ($\sim85$)
 satellites only 15 of which are known, or $\sim224$ ($\sim340$)
 across the sky. The ambiguity of Segue 1, Willman 1, and Bo\"otes II
 has considerable effect on the extrapolated MW census, underscoring
 the need for an understanding of these extremely faint systems.

These estimates assume that the sizes and luminosities of the known
satellites in DR6 are representative of the the MW satellite
population as a whole. Based on our detection limits we cannot make
any statements regarding extremely diffuse, low luminosity
systems that are undetectable by SDSS. The results also depend on the radial distribution
assumed. If we instead assume that the Milky Way's dwarf
population follows the radial distribution of the MW dSphs known prior
to 2004, then our $R_{90}$ inferred total number of dwarfs (with size-luminosities similar to known) within DR6 is 12, or 25 including  Segue 1, Willman 1 and Bo\"otes II. This implies that all or most satellites within DR6 would be known. 

\cite{tollerud} use the detection limits of \cite{koplf} to similarly estimate the true number of satellites within DR5 for a number of scenarios. The most comparable scenario to our assumptions (a limiting distance of 300 kpc, including all satellites except Segue 1 which is not in DR5) gives a result of $322^{+144}_{-75}$ satellites, consistent with our results of 232 for $R_{90}$ and 344 for $R_{99}$. 

%%%%%%%%%%%%%%%%%%%%%%%%%%%%%%%%%%%%%%%%%%%%%%%%%%%%%%%%%%%%%%%%%%%%%%%%%%%%%%%
\section{Allowing a Little Latitude}
Substantial effort has gone into the observation and interpretation of
the spatial distribution of the satellites of disk galaxies, in
particular that of the Milky Way satellites.  However, there is
neither agreement on whether the Milky Way satellites have a truly
anisotropic spatial distribution, nor whether we expect them
to. Pre-SDSS, \cite{Kroupa05} found that the distribution of known MW
satellites could be described by a disk of finite width, aligned
almost perpendicularly to the MW disk. This was in agreement with the
``Holmberg'' effect \citep{holmberg}, that the closest satellites to a
host galaxy were observed to be preferentially aligned with the minor
axis of the host. This disk-like distribution seemed incompatible with
$\Lambda$CDM, but \cite{Kang05} reasoned that if satellites follow the
distribution of the host dark matter profile rather than that of the
substructure then the dozen observed MW satellites could statistically
lie in a disk-like structure, although the orientation of this disk is
arbitrary. \cite{piatek} used proper motions derived from HST
observations to show that this ``Great Disk of Milky Way Satellites''
was not a persistent structure; the orbits of the dwarfs would not
contain them within this disk. \cite{metzpole} refute this conclusion,
finding instead that the orbital poles of most MW satellites place
them in a rotationally supported disk-of-satellites. Studies on the
satellites of other galaxies from SDSS also yield conflicting
results. \cite{bailin} affirm the Holmberg effect while
\cite{brainerd} find that satellites lie preferentially along the
major, not minor, axis of the host. \cite{zentner05} re-examine the
problem from a theoretical point of view, stating that DM substructure
is not completely isotropic and that the MW satellite distribution
can, albeit with a very low probability, be drawn from a DM subhalo
distribution. 

Within the standard $\Lambda$CDM structure formation scenario, satellite galaxies without dark matter could be formed
in gas-rich tidal tails during vigorous early galaxy--galaxy interactions \citep{tidaldwarfs}. Families of such tidal dwarfs would have correlated
orbital angular momenta and may appear as disk like arrangements about some hosts. This would support the apparent disk-of-satellites (Kroupa
et al. 2005; Metz et al. 2007) and its correlated orbital angular momenta (Metz et al. 2008).
It is therefore crucial to further constrain the spatial and orbital angular momentum properties of the satellites
to reveal their true nature which is intimately related to the formation of the MW.

\begin{figure}%[!ht]
 \center
 \includegraphics{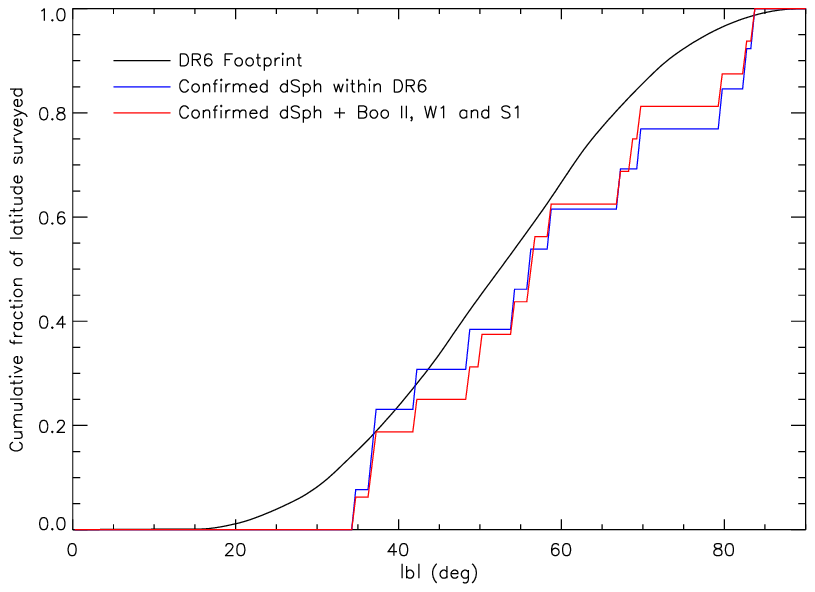}
 \includegraphics{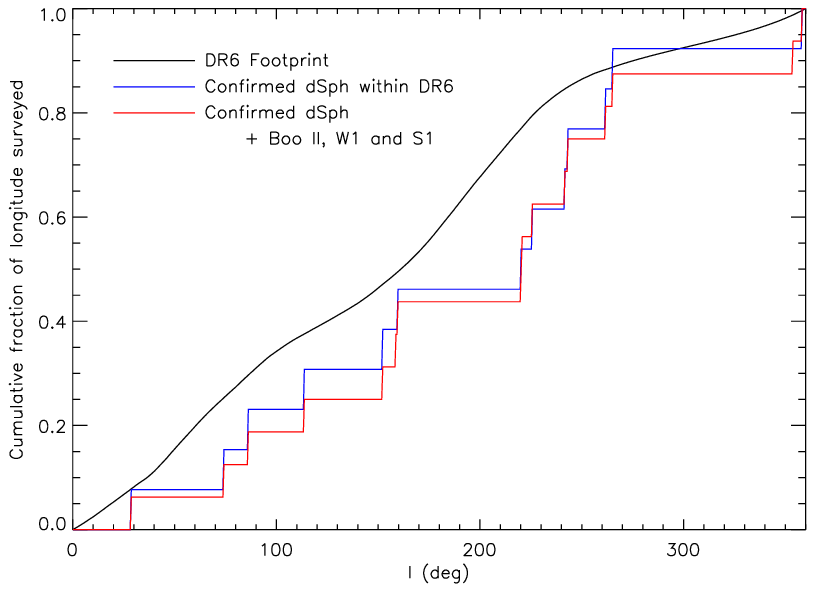}
 \caption{\emph{Top}: Cumulative histogram of the Galactic latitude of
 the DR6 footprint weighted by area (black). Blue shows the cumulative
 histogram of the latitudes of confirmed dSphs within DR6 and the MW
 virial radius (Boo, Dra, CVn, CVn II, ComBer, Leo I, Leo II, Leo IV,
 Her, UMa, and UMa II). Red shows the same but including Boo II,
 Willman 1, and Segue 1. \emph{Bottom}: Same as top panel, but for
 Galactic longitude.}
 \label{fig:latitude}
\end{figure}

A caveat of past studies of the MW dwarf distribution is that
the sky had not been uniformly searched for
satellites and the effect of Galactic latitude on the observability of dwarfs had not been thoroughly quantified. In their detailed study, \cite{kleynasearch} showed that latitude strongly affected the detectability of Milky Way satellites with their technique.  Our uniform study of SDSS DR6 takes their approach a step further and provides a detailed quantitative description of dwarf detectability over the footprint of our survey. We established in \S\ref{latitude} that the
average detectability of the known satellites does not significantly
vary over the DR6 footprint. We can thus compare the latitude and
longitude distribution of the Milky Way satellites within the DR6
footprint with that expected if they are randomly distributed.  We
perform a Kolmogorov-Smirnoff test to determine whether the satellites
detected in DR6 show statistically significant spatial
anisotropy. Figure \ref{fig:latitude} shows the cumulative
distribution (black lines) of latitude (top) and longitude (bottom) by
area of the DR6 footprint. Overplotted on both panels are the
cumulative distributions of MW satellites, both ignoring the ambiguous
objects Boo II, Willman 1 and Segue 1 (blue) and including them
(red). A KS test on these distributions with the entire DR6 area
yields a probability of isotropic distribution of 0.16 over longitude
and 0.72 over latitude, or 0.06 and 0.79 if we include Segue 1, Willman 1, and Bo\"otes II. We also randomly pick 12 (or 15) points from DR6
coverage, weighted by area, and repeat the KS test 1,000 times. The
mean resulting probabilities of isotropic distributions are
$0.45\pm0.29$ over longitude and $0.62\pm0.27$ over latitude, or
$0.34\pm0.28$ and $0.55\pm0.28$ with Segue 1, Willman 1, and Bo\"otes II. Hence, a conclusive result on the isotropy of the MW satellites
awaits further data as our test shows that considering DR6 alone
either scenario is plausible.

%%%%%%%%%%%%%%%%%%%%%%%%%%%%%%%%%%%%%%%%%%%%%%%%%%%%%%%%%%%%%%%%%%%%%%%%%%
\section{Towards SkyMapper}\label{SkyMapper}
SDSS Data Release 6 covers $\sim\pi$ steradians of Northern sky near
the North Galactic Pole, and at least nine new dSph companions have
been found in this area. The new ANU SkyMapper telescope will survey
the entire southern sky to a similar depth as SDSS over the next five
years, so we can naively expect to find around twenty-five new Milky
Way satellites. With a detailed, systematic search covering around
three quarters of the sky, we will for the first time be able to
conclusively compare the MW satellite galaxy population with
theoretical predictions. The apparent anisotropy of the satellites
will be conclusively confirmed or ruled out and we will continue to
discover the most dark matter dominated stellar systems nature
produced.
 
Designed as a replacement for the Great Melbourne Telescope that was
destroyed by the Canberra bush fires in January 2003, SkyMapper is a
new wide-field automated survey telescope currently being built at
Siding Spring Observatory, NSW. It will feature a 1.33m primary mirror
and a 0.69m secondary with an effective aperture of 1.13m and a $5.7$
square degree field of view. The imager consists of a $4\times8$
mosaic of $2048\times4096$ pixel CCDs \citep{SkyMapper}. SkyMapper's
primary purpose is the Southern Sky Survey which will be used to study
objects from Trans-Neptunian Objects to high redshift quasars. It's
five year mission: to survey the entire $2\pi$ steradians of the
southern sky at multiple epochs.
 
\cite{SkyMapper} states an average seeing at Siding Spring of
$\sim1.5$ arcsec as derived from observing logs of the
Anglo-Australian Telescope (AAT), on par with the median 1.4 arcsec
seeing of the SDSS site \citep{dr5}. The sky will be observed at six
epochs, and at completion 90\% of the sky will be observed at least
five times, and 100\% observed at least three times. The expected
survey depth (S/N = 5, t=110s per epoch) is given in Table
\ref{tbl:pars}. Also included are the corresponding magnitude limits
from SDSS \citep{dr5}. The survey aims to provide astrometry to better
than 50 mas, as compared with 100 mas for SDSS. Further information is
available at the SkyMapper
website\footnote{http://www.mso.anu.edu.au/SkyMapper}.
 
\begin{deluxetable}{cccc}
\tablecaption{Photometric depths of SkyMapper\tablenotemark{a} and
SDSS\tablenotemark{b} \label{tbl:pars}} \tablewidth{0pt} \tablehead{
\colhead{Filter} & \colhead{SM 1 epoch} & \colhead{SM 6 epoch} &
\colhead{SDSS}}
 \startdata \emph{u} & 21.5 & 22.9 & 22.0 \\
\emph{v}&21.3 &22.7 & - \\
\emph{g}&21.9 &22.9 &22.2 \\
\emph{r}&21.6 &22.6 &22.2 \\
\emph{i}&21.0 &22.0 &21.3 \\
\emph{z}&20.6 &21.5 &20.5 \\
 \enddata
 \tablenotetext{a}{Expected, \cite{SkyMapper}}
 \tablenotetext{b}{\cite{dr5}}
\end{deluxetable}

%%%%%%%%%%%%%%%%%%%%%%%%%%%%%%%%%%%%%%%%%%%%%%%%%%%%%%%%%%%%%%%%%%%%%%%
\section{Conclusion}
The dwarf galaxy satellites of the Milky Way provide excellent
opportunities to further our understanding of galaxy formation and
near-field cosmology. They can be resolved into individual stars
allowing detailed studies of their structure, kinematics and
composition. They have also been cause for concern regarding their
interpretation as the luminous components of dark matter substructure;
it has been argued that the number and spatial distribution of these
satellites are inconsistent with $\Lambda$CDM structure formation
scenarios. The commencement of the Sloan Digital Sky Survey triggered
a cascade of discoveries, with fourteen new satellites discovered. The
limited spatial coverage and photometric depth of SDSS suggests that
many, if not most, MW satellites are still yet to be discovered. The
coming years are likely to bring the MW satellite census towards
completeness as new survey telescopes such as SkyMapper, Pan-STARRS and LSST come online.

We present here the method used to search the Sloan Digital Sky Survey
Data Release 6 for ultra-faint Milky Way satellite galaxies. By
screening for stars consistent with an old population at a fixed
distance, we enhance the signal of a dSph over the Milky Way
foreground. Smoothing with a kernel corresponding to the expected
surface density profile further elevates the dSph above the
foreground, and our comprehensive thresholds account for varying
stellar density and more diffuse objects.

Applying our algorithm to SDSS DR6, we recover the ``classical'' and
recently discovered dSphs, as well as 17 globular clusters and two
open clusters. We also have 30 unidentified detections, some of which
may be new satellites. The discovery of Leo V demonstrates the difficulty in following-up dwarf candidates; while we detect Leo V, there are several unknown detections of greater significance that may prove to be something. However observing these weakest candidates is a rather hit and miss affair, as pointed out by \cite{leov}. 

To compare the known dwarf galaxy population of the Milky Way with
predictions, its essential to have a very well-defined dwarf selection
function.  To do this, we thoroughly model the detection efficiency of
systems covering a wide range in parameter space by simulating more
than 3,000,000 galaxies. We fit various functions to the resulting
detection efficiency contours to semi-analytically describe efficiency
as a function of magnitude, size, distance and Galactic
latitude. Using the results of our detailed investigation of dwarf
detectability, we show that:

\begin{itemize}
\item{Assuming a Via Lactea subhalo radial distribution and that $R_{complete}=R_{90}$, there
should be $\sim13$ satellites with DR6, 12 of which are known. If we
include Segue 1, Willman 1, and Bo\"otes II in this
calculation, this estimate jumps to $\sim56$, only 15 of which are
known.}

\item{Dwarf detectability shows a smooth transition from 100\% to 0\%
over size and distance. For example, the distance at which a CVn
II-like object is detected with 90\% efficiency is 200 kpc, compared
to 316 kpc for 10\% efficiency.}

\item{Galactic latitude does not significantly impact the detection of
satellites over the DR6 footprint, and surveys of similar quality
should still detect dwarfs as low as $b\approx15^{\circ}$. All of the
satellites discovered in SDSS would have been detected at any
latitude.}

\item{The census of MW satellites brighter than $M_V=-6.5$ should be
complete out to 300 kpc, and all objects brighter than $M_V=-5$ would
be detected with at least 50\% efficiency out to this distance.}

\item{Given the present data, the spatial anisotropy of the MW
satellites within DR6 cannot be confirmed or ruled out.}
\end{itemize}

We provide several different parameterizations of our detection limits
to facilitate comparisons between the known Milky Way dwarf galaxy
population and predictions.  We provide software that returns the
detection efficiency of a dwarf galaxy as a function of its
luminosity, scale size, distance, and latitude.  There are two
different codes provided for this; one is based on an analytic
description of our detection limits and the other provides a direct
interpolation from our 3,825,000 simulated galaxies.  We also provide
a linear fit as a function of $M_V$ of the distance out to which
dwarfs are detected with each of 50\%, 90\%, and 99\% efficiency.
These fits assume an underlying dwarf galaxy population with
combinations of sizes and luminosities similar to those known.

2009 will bring about the beginning of the Southern Sky Survey, and with it a way to uniformly search a further 20,000 square degrees of sky for new MW dwarfs. We will apply our algorithm to the incoming data and produce the most complete and well characterized census of the MW neighbourhood possible to date. We may also implement improvements to the algorithm to optimize for stellar streams or young stellar populations. \cite{lgtrio} have shown that M31 satellites occupy as yet unexplored size-luminosity space around the Milky Way. Surveys beyond SDSS and SkyMapper, such as Pan-STARRS and LSST, will be needed to carefully search for such systems.  Even with our carefully characterized detection limits the true number of MW satellites remains highly uncertain.

\acknowledgements We thank Bill Wyatt and the Telescope Data Center at
SAO for maintaining a copy of SDSS at the Harvard-Smithsonian Center
for Astrophysics.  SW thanks the Institute for Theory \& Computation
(ITC) at Harvard for support and hospitality and the Smithsonian
Astrophysical Observatory for financial support during the final
stages of this work. Thanks to Pavel Kroupa, Iskren Georgiev, Jose
Robles and Charley Lineweaver for helpful comments and discussion. HJ
and SW further acknowledge financial support from the Go8/DAAD -
Australia Germany Joint Research Co-operative Scheme and through the
Australian Research Council Discovery Project Grant DP0451426.

\end{document}